\documentclass{aa}  

\usepackage{graphicx}
\usepackage{multirow}
\usepackage[varg]{txfonts}
\usepackage{natbib}
\bibpunct{(}{)}{;}{a}{}{,}
\usepackage{soul}
\usepackage[dvipsnames]{xcolor}
\usepackage{threeparttable}
\usepackage{ulem}

\usepackage{siunitx}

\usepackage{booktabs, chemformula}

\begin{document} 

\title{Comprehensive laboratory constraints on thermal desorption of interstellar ice analogues}

\author{
    F.~Kruczkiewicz \inst{\ref{AMU}, \ref{LERMA}, \ref{MPE}},
    F.~Dulieu \inst{\ref{LERMA}},
    A.V.~Ivlev \inst{\ref{MPE}},
    P.~Caselli \inst{\ref{MPE}},
    B.M.~Giuliano\inst{\ref{MPE}},
    C.~Ceccarelli \inst{\ref{IPAG}}
    and P.~Theulé \inst{\ref{AMU}}
}


\institute{
Aix Marseille Univ, CNRS, CNES, LAM, Marseille, France 
\label{AMU}
\and
CY Cergy Paris Université, Observatoire de Paris, PSL University, Sorbonne Université, CNRS, LERMA, F-95000, Cergy, France
\label{LERMA}
\and
Max-Planck-Institut f\"ur Extraterrestrische Physik,
Gießenbachstraße 1, Garching, 85748, Germany
\label{MPE}
\and 
Université Grenoble Alpes, CNRS, IPAG, 38000 Grenoble, France 
\label{IPAG}
}

\date{Received --; accepted --}

\authorrunning{F.~Kruczkiewicz~et~al.}

\titlerunning{Laboratory constraints on ice sublimation}

 
  \abstract
   {Gas accretion and sublimation in various astrophysical conditions are crucial aspects of our understanding of the chemical evolution of the interstellar medium. To explain grain growth and destruction in warm media, ice mantle formation and sublimation in cold media, and gas line emission spectroscopy, astrochemical models must mimic the gas--solid abundance ratio. Ice-sublimation mechanisms determine the position of snow lines and the nature of gas emitted by and locked inside planetary bodies in star-forming regions. To interpret observations from the interplanetary and extragalactic interstellar mediums, gas phase abundances must be modelled correctly.}
    {We provide a collection of thermal desorption data for interstellar ice analogues, aiming to put constraints on the trapping efficiency of water ice, as well as data that can be used to evaluate astrochemical models. We conduct experiments on compact, amorphous H$_2$O films, involving pure ices as well as binary and ternary mixtures. By manipulating parameters in a controlled way, we generate a set of benchmarks to evaluate both the kinetics and thermodynamics in astrochemical models.}
   {We conducted temperature-programmed desorption experiments with increasing order of complexity of ice analogues of various chemical compositions and surface coverages using molecular beams in ultrahigh vacuum conditions (1 $\times$ 10$^{-10}$ hPa) and low temperatures (10 K). We provide TPD curves of pure ices made of Ar, CO, CO$_2$, NH$_3$, CH$_3$OH, H$_2$O, and NH$_4^+$HCOO$^-$, their binary ice mixtures with compact amorphous H$_2$O, ternary mixtures of H$_2$O :  CH$_3$OH : CO, and a water ice made in situ to investigate its trapping mechanisms.}
   {Each experiment includes the experimental parameters, ice desorption kinetics for pure species, and the desorption yield (gas--solid ratio) for ice mixtures. From the desorption yields, we find common trends in the trapping of molecules when their abundance is compared to water: compact amorphous water ices are capable of trapping up to 20\% of volatiles (Ar, CO, and CO$_2$), $\sim$ 3\% of CH$_3$OH, and $\sim$ 5\% NH$_3$ in relation to the water content within the ice matrix; ammonium formate is not trapped in the water ice films, and compact amorphous water ice formed in situ has similar trapping capabilities to a compact amorphous water ice deposited using molecular beams.}
   {Deposited or formed in a very compact structure, amorphous water ice of less than 100 layers cannot trap a large fraction of other gases, including CO and CO$_2$. These desorption yields offer insights into the availability of species that can react and form interstellar complex organic molecules during the warm-up phase of ice mantles. Furthermore, in order to be reliable, gas-grain astrochemical models should be able to reproduce the desorption kinetics and desorption yield presented in our benchmark laboratory experiments.}


   \maketitle
%

\section{Introduction}
\label{introduction}

Understanding the gas and dust cycle in the interstellar medium (ISM) is crucial for interpreting observations of both local and distant galaxies, as well as our own Galaxy and Solar System. Dust, which traps refractory elements (Fe, Si, Mg) in a solid state, condenses from gas in the dense outflows of dying low-mass stars and the remnants of type II supernovae \citep{NanniAA20}. In cold and dense molecular clouds, dust surfaces become covered with volatile compounds, forming icy mantles abundant in H$_2$O, CO, CO$_2$, CH$_3$OH, NH$_3$, and CH$_4$ \citep{BoogertARAA15, McClure_JWST}. These mantles are also chemically enriched by interstellar complex organic molecules (iCOMs), that is, C-bearing saturated molecules containing hetero-atoms and at least six atoms, such as methyl-formate (HCOOCH$_3$), dimethyl-ether (CH$_3$OCH$_3$), and acetaldehyde (CH$_3$CHO) \citep{Herbst2009ARA, Ceccarelli2017, Ceccarelli2022}. In the solid-phase, iCOMs are thought to form through both thermally activated, diffusive grain chemistry \citep{Garrod_Herbst2006, HerbstPCCP14} and non-diffusive reactions in dark cloud conditions \citep{TheuleASR2013, Fedoseev2015, Chuang2016, Garrod2022ApJS}.

From an astrochemical point of view, the sublimation of these ice mantles drives many processes in star-forming regions and planetary systems. For instance, in the early stages of star formation, thermal desorption dictates the so-called `hot corinos', which are regions around class 0 protostars that show bright emissions of gas-phase iCOMs thanks to the sublimation of water ices and these molecules as temperature rises above 100~K \citep{Cazaux2003, Ceccarelli2004, Chahine2022}. In later stages, protoplanetary discs exhibit temperature gradients that set the locations of snow lines, which are ice sublimation fronts that define the initial composition makeup of planets \citep{HarsonoAA15, MousisSSR20, Oberg_Bergin2021}. These snow lines are also crucial for understanding theories about the origins of water on Earth \citep{Trobin2023Natur}.

Although generic models often set the locations of major snow lines based on temperature gradients and characteristic sublimation temperatures of molecules using desorption parameters such as binding energies, in reality their location is influenced by various factors \citep{Oberg2023ARA}. These include dynamical effects in the temperature disc structure (such as outburst events) and chemical effects. On the chemical side, properties of the icy grains, such as the chemical composition of the ice, the grain surface, and the ice morphology (amorphous, crystalline, and porosity), also impact their sublimation temperatures. 

Experimental work on interstellar ice analogues has revealed that volatile species not only desorb over a range of temperatures but can also become trapped within diverse ice matrices. Studies have demonstrated that a variety of molecules, from hypervolatiles species (CO, N$_2$ and Ar) to H-bonding molecules (NH$_3$ and CH$_3$OH), can be effectively trapped  within H$_2$O-rich ices, with the entrapment efficiency being dependent on the deposition conditions and water-ice morphology  \citep{Collings_APJ03, Viti2004MNRAS, FayolleAA2011, MartinAA2014, BurkeBrown15, Lauck2015}. Similarly, CO$_2$-rich ices have been shown to serve as efficient matrices for trapping volatile species, which suggests the potential for a more complex distribution of volatiles within protoplanetary discs \citep{NinioGreenberg2017, Simon2019, Simon2023}. 

The presence of mixed-ice matrices therefore indicates that there may be multiple snow lines within a disc: the primary snow line at the expected sublimation point of pure ices, and secondary lines where trapped volatiles are released as the ice matrix transitions to a crystalline state and desorbs \citep{Oberg2023ARA}. In our Solar System, the entrapment of molecules within icy matrices provides critical insights into comet composition. The observed abundances of molecules such as CO, CO$_2$, N$_2$, and noble gases in cometary ices highlight the significance of entrapment processes within H$_2$O and CO$_2$ ice matrices, which is supported by both observational and experimental studies \citep{Kouchi1995, Villanueva2011, NinioGreenberg2017, Moussis2021}. In particular, recent findings for the comet 67P/Churyumov-Gerasimenko  from the ROSINA mass spectrometer reveal a close relationship between the densities of H$_2$O and CO$_2$ ices and the abundances of highly volatile molecules, such as CO and CH$_4$ \citep{Rubin2023}. This suggests that these volatiles were most likely trapped within the amorphous ices of the protoplanetary disc, long before the formation of the Solar System.  

Therefore, a deep understanding of thermal desorption processes in interstellar ice analogues is indispensable, and allows a more comprehensive view of the chemical parameters governing thermal desorption and is vital for refining gas-grain models that describe the gas-ice interplay during star formation. In the present study, we aim to provide a comprehensive set of laboratory data on interstellar ice analogues. Our primary goal is to offer data on the trapping efficiency of H$_2$O-rich ices, particularly their capacity to hold species. Our second goal is to furnish astrochemical models with data that can be used to benchmark them. For this, we propose a sequence of experimental tests ---specifically focused on thermal desorption and increasing in complexity--- that can be used to evaluate the balance between computational efficiency and the accuracy of gas-grain models.

The paper is structured as follows. Section~\ref{ice_parameters} outlines the ice desorption parameters we explored and Section~\ref{experimental} describes our experimental setup and methodology. The results, which provide a guide for future model benchmarking, are presented with progressively increasing complexity in Sections~\ref{results_kin} and \ref{results_yields}. In Section \ref{results_kin}, we introduce the desorption of pure solids under our experimental conditions, arranging the components in ascending order of desorption energy as categorised by \cite{Collings_2004MNRAS_survey} and \cite{Viti2004MNRAS}: Ar, CO, CO$_2$, NH$_3$, CH$_3$OH, H$_2$O, and NH$_4^+$HCOO$^-$. Subsequently, we describe the desorption of binary ice mixtures, again sorting the species in increasing order of desorption energy: CO:H$_2$O, Ar:H$_2$O, CO$_2$:H$_2$O, NH$_3$:H$_2$O, CH$_3$OH:H$_2$O, and NH$_4^+$HCOO$^-$:H$_2$O. Lastly, we study the ternary ice mixtures H$_2$O:CH$_3$OH:CO. In section \ref{results_yields}, we focus on the desorption yields (ice trapping) for each desorption component. We place particular emphasis on CO in water ice, as CO is the most abundant and significant molecule after H$_2$ and the most challenging to investigate. We also systematically vary the initial morphology of the water substrate, either by creating a compact ice using a molecular beam or by generating the H$_2$O molecule in situ through the O$_2$ + H + H reaction \citep{Accolla_MNRAS_2013}. The key experimental findings are discussed in Section \ref{Discussion}. In the final Section \ref{conclusions}, we present the astrophysical significance of our findings and propose primary guidelines for gas-grain model benchmarking based on our results. We also provide numerous references to detailed studies in the literature for each topic we cover. 

\section{Thermal desorption parameters and gas-grain modelling}
\label{ice_parameters}

The thermal desorption process is influenced by a variety of parameters, including the chemical composition, surface coverage, morphology, and heating rate of the ice. In this section, we describe these parameters and discuss their importance in the context of interstellar ice sublimation processes:

\begin{itemize}
    \item \textbf{Chemical composition}: The desorption kinetics of a molecule is well accounted for by the Wigner-Polanyi equation and has a desorption rate, which is the inverse of the residence time on the grain \((k_{des} = \tau ^{-1} = \nu \cdot exp(-E_{bind}/RT)\), which is accounted by two parameters: the binding energy \(E_{bind}\) and the pre-exponential factor \(\nu\). They both depend on the chemical nature of both the adsorbate and the surface. Molecules can be conveniently classified by increasing the order of the binding energy: CO-like molecules, intermediate molecules, H$_2$O-like molecules, and semi-volatile molecules \citep{Collings_2004MNRAS_survey}. Refractory molecules typically involve a covalent bond. They chemisorb at temperatures exceeding 400 K. It is important to note that the binding energy of the same molecule can vary, it changes depending on the surface \citep{Collings_ASS03, NobleAA12} and the neighboring species it is interacting with \citep{NguyenAA18}. In fact, a molecule might show a range of desorption energies \citep{Dulieu2005, FerreroApJ2020, Minissale_review_22}. An important detail to consider is the desorption process. If it is not elementary, the pre-exponential factor, $\nu$(T), might vary with temperature. This accounts for the multiple stages involved in desorption.
    
    \item \textbf{Ice thickness}: If the 2D desorption kinetics from a grain surface is well accounted for by the Wigner-Polanyi equation and totally desorb, the 3D desorption of molecules embedded in a water ice mantle exhibits additional desorption features related to water ice crystallisation \citep{JenniskensScience94, KayPRL97} and co-desorption \citep{Sandford_Allamandola_icarus_88, Collings_2004MNRAS_survey} with respective desorption yields \citep{Viti2004MNRAS, MayI_JCPA_2013, MayII_JCPA_2013, MartinAA2014}. Indeed, the desorption yields are as important as their desorption kinetics since they set both the amount of released gas available for warm gas-phase chemistry and the amount of remaining material on grains that will be later incorporated in planetesimals and comets. 
    
    \item \textbf{Ice morphology}: Water is a highly polytropic material. The exact morphology of interstellar ices is still an open question, and it is not yet established if they are porous or compact. In laboratory settings, a film of amorphous ice will form at low temperatures, if pores are present they will collapse over 40 K,  while at higher temperatures (above 120 K) amorphous ice can become crystalline. This morphology changes affect the effective surface area at play during percolation, the desorption \citep{MayI_JCPA_2013, MayII_JCPA_2013} and reaction-diffusion processes \citep{GhesquiereAA18}.
    
    \item \textbf{Heating rate}: In a linear temperature ramp, T(t) = T$_0$ + $\beta~\cdot$ t, the inverse of the heating rate $\beta$ [K.s$^{-1}$], is the time spent on each K. Since desorption and crystallisation are activated time-dependent processes, they depend on the coupling of time and temperature through the $\beta$ [K.s$^{-1}$] heating rate parameter. It is an important parameter to transpose laboratory experiments to astrophysical environments, as they have very different timescales.
    \end{itemize}

To account for these accretion/desorption processes several gas-grain models were developed. They can be classified into several categories:

\begin{itemize}
\item Two-phase models: these early models \citep{TielensAA82,dHendecourtAA85,HasegawaApJS92} treat the grains as an infinite surface where all the molecules desorb in the sub-monolayer regime. This rate equation formalism is simple, robust and useful in many situations. Later refinements take into account the discrete aspects of the grains \citep{BihamApJ01, GreenAA01,GarrodAAP08}. This formalism is adopted by major codes, such as PDR Meudon \citep{LePetitApJS06}, CLOUDY \citep{FerlandRMxAA17}, MAPPINGS \citep{SutherlandASCL18} and many others where grain chemistry is not central. 
\item Three-phase models: they are an extension of the 2-phase models, the distinction between the surface and the mantle is made through a core coverage factor \citep{HasegawaMNRAS93}. These modified rate equation models can treat the mantle with little computational effort. This is the case of NAUTILUS \citep{RuaudMNRAS16}, UCLCHEM \citep{HoldshipAJ17}, or MAGICKAL \citep{GarrodApJ11}, and GRAINOBLE \cite{TaquetAA12} and other codes.    
\item 3D models: these codes treat precisely the multilayer aspect of the ice, determining the position of the particles explicitly and often using a Monte-Carlo approach to treat diffusion \citep{ChangAA07,VasyuninApJ13, GarrodApJ13}. 3D models are computationally more expensive, but they better represent the formation of interstellar complex organic compounds in ice. \end{itemize}

These models correspond to different needs and balances in terms of precision, computational power and speed. They rely on laboratory work \citep[e.g.][]{Collings_2004MNRAS_survey} or quantum chemistry calculations \citep[e.g.][]{Ferrero2022}, both for the numerical values and for the physical description they use. Consequently, they should be capable, in principle, to model back the laboratory experiments from which they derive their data. 

This paper aims to explore the four-parameter ice desorption parameter space, offering a more extensive and systematic approach than previously attempted. While not exhaustive, this paper presents a complete and incremental benchmarking set, with all necessary information readily accessible. Experiments were mainly conducted on a compact amorphous H$_2$O ice matrix to investigate how the entrapment of molecules occurs if the porosity is significantly reduced. To the best of our knowledge, entrapment of astrophysical ice analogues was investigated experimentally only in highly porous ice matrices \citep[e.g.][]{Collings_2004MNRAS_survey, FayolleAA2011, MartinAA2014}, while \cite{MayI_JCPA_2013, MayII_JCPA_2013} studied the crack propagation of compact ices. It is not yet really established if interstellar ices are porous or not, but we just recall here that porosity is due to the mode of deposition of water molecules from the gas phase to the solid phase \citep{Stevenson1999, KimmelJChPh2001_exp}. However, compact ice is more stable than porous ice, and the latter tends to compact with time and temperature. Moreover, compaction is induced by energetic processes such as chemical reactions \citep{Accolla2011}, UV radiation, or irradiation simulating cosmic rays \citep{MEJIA2015}. Finally, if water ice is formed on the grains, it is compact by nature \citep{Accolla_MNRAS_2013}. Therefore, experiments using compact ice films, even if they have not been used much in the past, are of paramount importance.

For each experiment, we give the necessary input parameters and derive as output both the kinetics of the desorption and the various desorption yields, which will serve as benchmarks for astrochemical models. Several simplifications, which are indicated below, are made in data analysis as a compromise between precise physical treatment and convenient gas-grain modelling. 

Indeed the kinetics of desorption is only one face of the problem. A description solely based on the Wigner-Polanyi rate equation for desorption implicitly assumes complete desorption of the material. The other face of the problem is the amount of gases released in each peak of the multi-desorption peak pattern. While this aspect is as important as the kinetic aspect, it is much less explored and most of the time not considered in gas-grain models. Actually, if most of the models have a rather correct description of the surface species desorption kinetics, modelling the diffusion-desorption and desorption yield of bulk species is more difficult, and this has been the subject of much fewer laboratory studies \citep{Collings_2004MNRAS_survey, BolinaSurSc05, FayolleAA2011, MayI_JCPA_2013, MayII_JCPA_2013, MartinAA2014, Cooke_APJ_18, Simon2019, Simon2023}. In this work, we give the desorption yields for all the experiments performed. Besides the out-of-equilibrium description of gas desorption offered by the Wigner-Polanyi equation, sublimation is also commonly described using an equilibrium description based on vapour pressures \citep{Schmitt92}. Nevertheless, a comparison between both approaches and a discussion of their limits of applicability are out of the scope of this paper. Here, we focus on the Wigner-Polanyi description, which is the most commonly used in ISM studies.

\section{Methods}
\label{experimental}

\subsection{Experimental setup}

\begin{figure*}[ht]
   \centering
   \includegraphics[width = 15 cm]{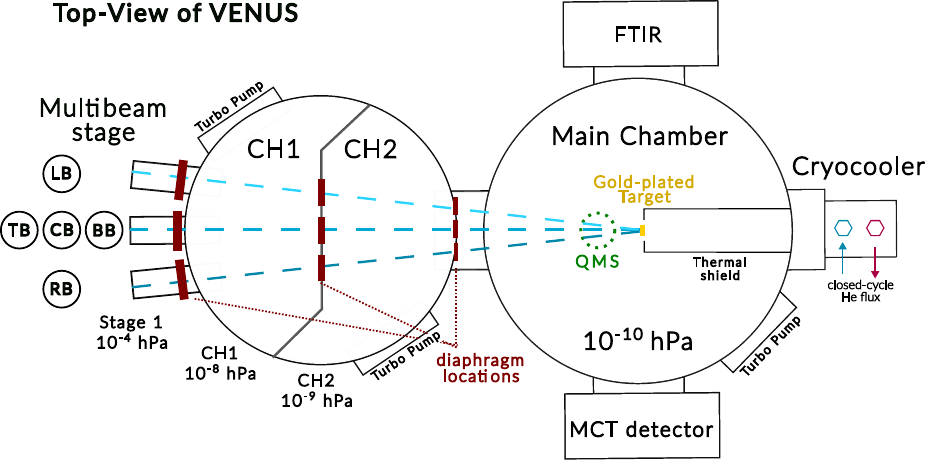}
   \caption{
Schematic of the VENUS setup viewed from above, showing a multi-beam stage with five beam lines (Left, Top, Central, Bottom, and Right beams) for material injection, symmetrically distributed around the central port. The CB and BB are vertically aligned below the TB. Each beam is equipped with a set of diaphragms to collimate the particle stream into the vacuum chambers CH1 and CH2 at pressures of 10$^{-8}$ hPa and 10$^{-9}$ hPa, respectively. The main chamber (10$^{-10}$ hPa) houses a gold-plated target for growing the ice samples. Gas-phase species are monitored by a QMS that is placed in front of the sample during TPD analysis. A cryocooler and thermal shield regulate the system's temperature.}
   \label{fig:venus_diagram}
    \end{figure*}

The results presented in this paper were obtained using the multi-beam ultrahigh vacuum (UHV) apparatus named VENUS (VErs les NoUvelles Synthèses - Towards New Synthesis), housed at the LERMA laboratory in CY Cergy Paris Université. VENUS simulates the solid-state, non-energetic formation conditions of interstellar complex organic molecules (iCOMs) in dark molecular clouds and circumstellar environments. A detailed description of its technical aspects, as well as full test calibration results to understand the set-up capabilities, is provided by \cite{Congiu_RSI_2020}. The essential features of the VENUS apparatus relevant to this study are described below and a schematic representation of the setup is provided in Fig. \ref{fig:venus_diagram}.

The VENUS setup includes two high vacuum chambers, CH1 and CH2, which form a compact two-stage differentially pumped path for the beam lines, and a UHV chamber. The UHV chamber, with a base pressure of 1~$\times$~10$^{-10}$~hPa at 10K, serves as the main reaction chamber. 

The sample holder, used for ice growth, consists of a 1 cm diameter circular copper mirror coated with gold and it is attached to the cold head of a closed-cycle He cryostat. The surface temperature of the sample is monitored using a silicon diode sensor and controlled within a range of 8 K – 400 K by a Lakeshore 340 controller, achieving a stability of $\pm$ 0.1 K and an accuracy of $\pm$ 2 K.

Gas introduction into VENUS occurs through a 4 mm inner diameter tube prior to the first chamber, producing an effusive beam. This beam traverses three chambers (stage 1, CH1, and CH2) before entering the main chamber via three diaphragms. A crucial aspect of the VENUS set-up is the alignment of the beam line, which involves the source (a 25 cm long quartz tube), the three diaphragms, and the target. This alignment, verified by a laser beam, ensures all beams converge precisely on the same spot on the sample, maximising the flux intensity and physical-chemical interactions of different species on the surface deposited through the beam lines. The local equivalent pressure in the beam volume is about 10$^{-8}$ hPa, which makes any gas phase reactivity or interactions negligible.

For gas-phase species analysis, VENUS employs a Hiden 51/3F quadrupole mass spectrometer (QMS), positioned 5 mm from the sample. The QMS can be adjusted vertically and rotated to suit different measurement needs. In its lower position, it analyses residual gas in the main chamber. For Temperature-Programmed Desorption (TPD) analysis, the QMS is raised and aligned with the sample. The QMS can also measure the flux and the composition of each beam. The ionising electrons in the QMS are set at 30 eV to minimise molecule fragmentation, ensuring accurate gas-phase analysis.

\subsection{Experimental procedure}

The preparation of ice analogues involves condensing gases from well-collimated beams onto a gold-coated substrate, typically at a deposition temperature of 10 K. The use of well-collimated beams with a normal incidence aligns with findings from \cite{Stevenson1999, KimmelJChPh2001_exp} that indicate water ices formed under these conditions are amorphous and compact.

In our experiments, we employ a compact amorphous H$_2$O ice matrix to study trapping mechanisms under reduced porosity conditions. The preference for compact over porous ice films lies in their higher reproducibility, achieved through a more uniform distribution of binding sites and a reduced likelihood of pore collapse, which can affect reorganization kinetics during thermal processing.

VENUS is equipped with five molecular/atomic beam lines (and four currently operational: Top, Central, Bottom and Right), enabling the simultaneous introduction of various atomic or molecular species onto the gold-coated surface. Liquid samples, such as water or methanol, are introduced through the central beam line and controlled by a needle valve. Gaseous species, on the other hand, enter through the top beam line, with their flow regulated by an automated Bronkhorst High-Tech control valve (1 SCCM\footnote{Standard Cubic Centimetre per Minute: 1 SCCM = 592 m$^3$ Pa s$^{-1}$ in SI units.}). This automated system ensures consistent molecular flow, vital for reproducibility across experiments conducted on different days. The gases, after passing through three stages of differential pumping and two diaphragms, form a well-collimated beam (2-mm diameter) that reaches the main chamber with an incremental pressure increase of less than 1 $\times$ 10$^{-10}$ hPa for gases with a molar mass larger than 16 (e.g, not H, H$_2$ nor He).

For binary and ternary ice mixture experiments, the central beam line is utilised for depositing water ice or mixtures of water and methanol, while gaseous species like CO are introduced via the top beam line. The right beam line houses a microwave discharge to generate atomic hydrogen 
by the dissociation of H$_2$ molecules. The flow rate of the gaseous species is adjustable and set to specific values (0.5, 0.3, or 0.1 SCCM) to achieve the desired mixture ratio.

Post-deposition, sample analysis is conducted using the TPD technique. This involves a gradual increase in surface temperature at a controlled rate, alongside monitoring the desorption signal from the mass spectrometer in relation to the sample's temperature.

\subsection{Surface coverages}

In our experiments, ice mixtures are deposited with a homogeneous distribution of components using VENUS different beam lines concurrently and the deposition doses are quantified in monolayer (ML) units, where 1 ML is approximated as 1 $\times$ 10$^{15}$ molecules cm$^{-2}$. This value represents the number of adsorption sites per cm$^2$ on compact amorphous solid water (c-ASW) and serves as a unit of surface density. For a fully wetting molecule like CO or Ar, a coverage of 1 ML indicates that all binding sites are occupied on the ice surface. Any additional molecules would then adsorb in new binding sites on top of the adsorbed molecules and not in the substrate, initiating the formation of a second monolayer. In contrast, on porous substrates, the number of binding sites per cm$^2$ increases with the thickness of the water ice film, given the 3D structure of porous water ice. Therefore, a larger deposition dose is required to actually enter the physical second layer regime, while the deposition dose is larger than one ML. Some molecules like CO$_2$ do not wet the surface and form islands. Therefore it is important to note that this corresponds to a surface density unit, that is used as a easy way to control the amount of molecules on the surface, but does not necessarily corresponds to a physical layer.

The interaction dynamics vary between molecules adsorbed in the substrate and those adsorbed on top of other molecules. Molecules in the first monolayer exhibit stronger substrate interactions, while those deposited on top of them primarily demonstrate molecule-molecule interactions due to weaker contact with the substrate. This distinction is crucial as it allows us to define the surface coverage of sub-monolayer to monolayer regime. 

To ascertain the surface coverage of each species, we utilize the TPD curves and a calibration method. The total number of molecules desorbed is proportional to the integrated ion current of a specific mass-to-charge (m/z) ratio recorded by the QMS. We determine the ice coverage by comparing the relative integrated areas of the TPD curves to the known area for 1 ML of CO. This benchmark was established via "family of TPDs" experiments (see Fig. \ref{CO_familly} in the Appendix), where CO was incrementally deposited onto c-ASW, enabling the identification of the dose equivalent to 1~ML. This approach measures the actual amount of species deposited on the target, rather than just the beam flux, revealing the "filling behaviour" transition from sub-monolayer to monolayer regime \citep{KimmelJChPh2001_exp, NguyenAA18}.

For species where this sub-monolayer to monolayer transition does not manifest via a `filling behaviour', we employ a modified procedure based on \cite{Martin2015}. The abundance of a species \( N_X \), derived from its integrated QMS signal area (\( A\)), is calibrated against that of CO using the relation:

\begin{equation}
N_{mol} = {N_{CO}} \frac{A(m/z)}{A(28)} \cdot \frac{\sigma^+(CO)}{\sigma^+(mol)} \cdot \frac{I_f([CO]^+)}{I_f([z]^+)} \cdot \frac{F_f(28)}{F_f(m)} \cdot \frac{S(28)}{S(m/z)}, 
\label{N_X}
\end{equation}  

\noindent here, \( A(m/z) \) is the integrated area under the QMS signal for the specific mass-to-charge ratio (\( m/z \)), and is directly linked to the abundance of the species. \( N(CO) \) is the surface coverage of CO, obtained from its desorption profile in the "family of TPDs" experiments, with \( A(28) \) being the respective curve area. The ionisation cross-section \( \sigma^+(mol) \), set at an electron impact energy of 30 eV, corresponds to the ionisation efficiency. The ionisation factor, \( I_F(z) \), is assumed to be one with no double ionisation of the molecules taking place in the QMS. \( F_F(m) \) is the fragmentation factor, indicating the likelihood of molecular fragmentation and it is inferred from the QMS data acquired during the TPD experiments. Lastly, \( S(m/z) \) is the sensitivity of the QMS at a specific mass. The parameters used in Eq. \ref{N_X} are detailed in Table \ref{paramenters_QMS}.
 
\begin{table}[h]
\centering
\begin{threeparttable}
\small
\caption{List of parameters used in Eq. \ref{N_X} to obtain the surface coverage for a given species.}
\begin{tabular}{lccccc}
\hline
Molecule & \( \sigma^+(m/z) \)\tnote{a} & \( F_F(m/z) \) & \( I_F(z) \) & \( m/z \)& \( S(m/z) \)\tnote{b} \\ \hline
CO         & 1.296\tnote{c}                                                                  & 1                   & 1                  & 28              & 1                  \\
Ar         & 1.840\tnote{d}                                                                   & 1                   & 1                  & 40              & 0.98               \\
CO$_2$ & 1.698\tnote{c}                                                                  & 0.95                & 1                  & 44              & 0.80               \\
NH$_3$        & 1.909\tnote{e}                                                                  & 0.55                & 1                  & 18              & 1.16               \\
CH$_3$OH      & 3.446\tnote{f}                                                                  & 0.72                & 1                  & 32              & 0.95               \\
H$_2$O        & 1.262\tnote{c}                                                                  & 0.90                & 1                  & 18              & 1.16               \\
NH$_4$$^+$HCOO$^-$    & 2.150\tnote{g}                                                                  & 0.55                & 1                  & 64              & 0.54               \\ \hline
\textbf{Notes. }
\end{tabular}
\begin{tablenotes}
   \item[a] Ionisation cross section from electron impact of 30 eV, units of 10$^{-16}$ cm$^{-2}$. \tnote{c}\cite{orient1987electron}. \tnote{d}\cite{straub1995absolute}. \tnote{e}\cite{Itikawa2017}. \tnote{f}\cite{NIXON201648}. \tnote{g} As there are no value for ammonium formate in the literature, we assumed the one from formic acid reported from \cite{Zawadzki2018}.
   \item[b] Sensitivity of the QMS obtained from \cite{HidenAnalytical2023}.
\end{tablenotes}
\label{paramenters_QMS}
\end{threeparttable}
\end{table}

Calibration experiments were conducted to achieve the desired ice composition. For instance, the injected pressure flux for water was adjusted to form 1 ML every 9 minutes. To maintain a 2 : 1 deposition ratio with CO, the automated flow regulator was set to form 1 ML of CO over an 18-minute interval (0.5 SCCM). This preparatory step is fundamental, as the surface coverage of the samples is ultimately deduced from the final TPD spectra, accommodating minor fluctuations in the actual deposited ratios.

The uncertainty in the surface coverage of CO and Ar is dominated by the error in estimating 1 ML from TPD curves in time steps, CO $\approx$ 10\% and Ar $\approx$ 15\%. For other molecules, our uncertainty analysis faces an additional challenge due to the lack of direct calibration data for our QMS. To address this, we have adopted sensitivity values from a similar QMS reported in the literature. However, it is important to note that this approach introduces an additional layer of uncertainty, as the calibration characteristics of different QMS instruments can vary. Considering all sources of errors, a 25\% uncertainty on calculated surface coverages is likely conservative enough.

\subsection{Species selection}
\label{species_selection}

Molecules can be categorised as either closed-shell species, meaning they have completely filled electron shells, or open-shell species, commonly known as radicals. Although radicals are crucial in interstellar solid-state reactions, their direct laboratory characterisation poses challenges. For the scope of this paper and due to these experimental constraints, our study focuses on closed-shell species. In alignment with \cite{Collings_2004MNRAS_survey} and \cite{Viti2004MNRAS}, we classify the species into five distinct categories based on their desorption behaviour in homogeneous binary ice mixtures with water.

\textit{i. CO-like species:} This category comprises hyper-volatile molecules such as CO, Ar, N$_2$, O$_2$, and CH$_4$. These species exhibit a desorption peak at low temperatures (below 60 K) corresponding to their pure desorption. Additionally, they can be trapped within water ice and subsequently desorb either during the amorphous-to-crystalline water transition — known as the 'volcano peak' — or through co-desorption with crystalline water. In this study, we specifically focus on CO and Ar. 

\textit{ii. Intermediate species:} Molecules like CO$_2$, H$_2$S, HCN, OCS, C$_2$H$_2$, SO$_2$, CS$_2$, and CH$_3$CN exhibit higher binding energies that fall between those of CO-like and H$_2$O-like species. For this category, CO$_2$ was chosen for study.

\textit{iii. H$_2$O-like species:} When a molecule can form hydrogen bonds with water, the interaction between water and that species intensifies. Such molecules, with shifted binding energies closer to those of water ice, include NH$_3$, CH$_3$OH, HCOOH, and CH$_3$CH$_2$OH. In this category, we focus on NH$_3$ and CH$_3$OH.

\textit{iv. Semi-volatile/semi-refractory species:} These are compounds that sublimate at temperatures above that of water but below that of refractory materials, such as minerals. Examples include some ammonium salts, represented as NH$_4$$^+$A$^-$, where A$^-$ is an anion such as formate (HCOO$^-$), hydrosulfide (SH$^-$), acetate (CH$_3$COO$^-$), cyanate (OCN$^-$), or hydrogen cyanide (HCN$^-$); sugars, like glycolaldehyde (CH$_2$OHCHO); and amino acids, such as glycine (NH$_2$CH$_2$COOH). For this study, we have selected the ammonium formate (NH$_4$$^+$HCOO$^-$) salt for investigation.

\textit{v. Refractory Species:} These are species that desorb at high temperatures (above 400~K, reaching up to thousands of Kelvins) and usually comprise minerals forming the core of the grain, such as SiO$_2$ and Mg. These species are not addressed in our study.

It is important to note that in our experiments, species were always mixed homogeneously; layering was not considered. While layering can be a factor such as in the case of CO freeze-out on top of the icy mantle in prestellar cores, our primary aim is to provide experimental data suitable for astrochemical models. Thus, benchmarking and trapping values can be based on these homogeneous mixtures.

\subsection{Summary of the experiments}

An overview of the experiments is provided in Table \ref{table_overview}, and the TPD data relevant to this study are available in our online repository\footnote{Data can be accessed at \url{https://lerma.labo.cyu.fr/DR/public-data.html}}. The heating ramp was set at 0.2 K s$^{-1}$ for all but one experiment (Exp. 23). The notation ${\text{sample A + sample B}}_{A : B}^T$ is used throughout this paper to indicate that the gases were co-deposited at temperature T, and the estimated final doses for each sample A : B are given in ML. If the sample is compact, one can estimate the thickness by knowing that one ML corresponds to 0.3 nm. Of course, the total thickness is the sum of individual thickness so doses (in the case of compact layers adapted to our beam deposition method). 
For example, $\{H_2O + CO\}_{2 : 1 ML}^{10K}$ signifies the co-deposition of 2 ML of H$_2$O and 1 ML of CO at 10 K onto the gold-coated surface. The estimated thickness is around 0.9 nm. According to existing literature, the average H$_2$O ice coverage in the ISM is estimated to range from a sub-monolayer up to 30 ML \citep{Potapov2020}, and possibly even a few hundred ML according to some sources \citep{Dartois2005}. In our experiments, the final doses did not exceed 85 ML, and most had a final dose below 30 ML, offering a reasonable approximation of ISM conditions.

\begin{table*}[ht]
\caption{Experiments overview and corresponding desorption yields.}
\label{table_overview}
\centering
\begin{threeparttable}
\begin{tabular}{
    l
    c
    c
    S[table-format=2.2]
    S[table-format=2.2]
    S[table-format=2.2]
    S[table-format=1.2]
    S[table-format=1.2]
    S[table-format=1.2]
    S[table-format=2.0]
}
\toprule
Exp & {Molecule} & {Ratio} & {H$_2$O} & {\(x\)}  & {Final Cover.} & {Beta} & {\(x_{\text{trapp}}\)} & {\(x_{\text{eff-water}}\)} & {\(T_{\text{dep}}\)} \\
& \(x\)  & {H$_2$O : \(x\)} & {ML}  & {ML} & {ML} & {K.s\(^{-1}\)} & {} & {} & {} \\ 
\midrule
\multicolumn{10}{c}{pure ice} \\
\midrule
\addlinespace
1   & H$_2$O  & 1 : 0 & { } & 21.5  & 21.5 & 0.2  & 0     & 0    & 10   \\
2   & CO       & 0 : 1 & { } & 1.6  & 1.6  & 0.2  & 0     & 0    & 10   \\
3  & Ar       & 0 : 1 &  { }  & 3.0  & 3.0      & 0.2  & 0     & 0    & 10   \\
4 & CO$_2$ & 0 : 1 &   { }   & 2.4  & 2.4      & 0.2  & 0     & 0    & 10   \\
5 & NH$_3$       & 0 : 1 & { }    & 3.0  & 3.0      & 0.2  & 0     & 0    & 10   \\
6  & CH$_3$OH & 0 : 1 & { }  & 14.7 & 14.7     & 0.2  & 0     & 0    & 10 \\
7   & NH$_4$$^+$HCOO$^-$ & 0 : 1 & { }  & 3.5  & 3.5      & 0.2  & 0     & 0    & 10   \\
\addlinespace
\midrule
\multicolumn{10}{c}{binary ice mixtures} \\
\midrule
8   & CO  & 2 : 1 & 1.82  & 0.93  & 2.75 & 0.2  & 0     & 0    & 10   \\
9  &     &       & 3.55  & 1.70  & 5.25      &      & 0.06  & 0.03 &      \\
10   &     &       & 6.55  & 3.70  & 10.25     &      & 0.11  & 0.06 &      \\
11   &     &       & 9.20  & 4.55  & 13.75     &      & 0.22  & 0.11 &      \\
12  &     &       & 15.44 & 6.96  & 22.40     &      & 0.30  & 0.14 &      \\
13 &     &       & 20.95 & 11.65 & 32.60     &      & 0.27  & 0.15 &      \\
14   &     &       & 40.50 & 19.10 & 59.60     &      & 0.34  & 0.16 &      \\
15    &     & 3 : 1 & 3.36  & 1.04  & 4.40      & 0.2  & 0.07  & 0.02 & 10   \\
16   &     &       & 6.00  & 2.00  & 8.00      &      & 0.17  & 0.06 &      \\
17  &     &       & 9.45  & 3.20  & 12.65     &      & 0.25  & 0.08 &      \\
18 &     &       & 19.70 & 6.95  & 26.65     &      & 0.35  & 0.12 &      \\
19  &     &       & 36.55 & 12.25 & 48.80     &      & 0.48  & 0.16 &      \\
20   &     &       & 61.70 & 21.15 & 82.85     &      & 0.53  & 0.18 &      \\
21  &     & 5 : 7 & 8.45  & 11.85 & 20.30     & 0.2  & 0.04  & 0.05 & 10   \\
22 &     &       & 16.77 & 22.05 & 38.82     &      & 0.06  & 0.08 &      \\
23 &     & 3 : 1 & 17.40 & 6.10  & 23.50     & 0.02 & 0.28  & 0.10 & 10   \\
24  & Ar  & 2 : 1 & 6.12  & 3.14  & 9.26      & 0.2  & 0.10  & 0.05 & 10   \\
25   & CO$_2$& 2 : 1 & 10.55  & 4.40  & 14.95  & 0.2  & 0.19  & 0.08 & 10   \\
26  &     &        & 23.55 & 9.45  & 33.00  &        & 0.31  & 0.12 & 10   \\
27 & CH$_3$OH    &      & 8.20   & 1.27  & 9.47   & 0.2    & 0.29  & 0.034 & 10   \\
28 &             &        & 13.84 & 0.54  & 14.38  &        & 1.52  & 0.028 & 90   \\
29 &             &        & 12.09 & 2.34  & 14.46  &        & 0.19  & 0.031 & 90   \\
30 &             &        & 19.53 & 0.97  & 20.50  &        & 1.43  & 0.033 & 90   \\
31 & NH$_3$    & 2 : 1  & 4.02  & 2.20  & 6.22   & 0.2    & 0.08  & 0.044 & 10   \\
 32  & NH$_4$$^+$HCOO$^-$ &        & 15.0 & 3.5 & 18.7  &  0.2   & 0 & 0 & 10   \\
\addlinespace
\midrule
\multicolumn{10}{c}{ternary ice mixtures} \\
\midrule
33 & CH\(_3\)OH : CO &  2 : 1\tnote{a}  & 8.97  & 5.02  & 14.00  & 0.2  & 0.08  & 0.05 & 10  \\
34  &     &  2 : 1\tnote{a}  & 15.79 & 9.63  & 25.42  &      & 0.16  & 0.10 & 10   \\
\addlinespace
\midrule
\multicolumn{10}{c}{water ice formed \textit{in situ}} \\
\midrule
 35  & CH$_3$OH : CO : CO$_2$ &  2 : 1\tnote{b} & 9.56 & 4.07  & 13.63  & 0.2  & 0.13  & 0.05 & 10 \\
 \addlinespace
\bottomrule
\end{tabular}
\begin{tablenotes}
   \item[a] These values correspond to the ratio for H\(_2\)O : CO in the ternary mixture with methanol.
   \item[b] These values correspond to the ratio for H\(_2\)O : CO in the mixture of water formed \textit{in situ}.
\end{tablenotes}
\end{threeparttable}
\end{table*}

\section{Results: Kinetic parameters}   
\label{results_kin}

We determined the binding energies for multi-layer coverages of Ar, CO, CO$_2$, NH$_3$, CH$_3$OH, H$_2$O and NH$_4$$^+$HCOO$^-$ using the approach described below in section \ref{kinetic_param}. The kinetic parameters for their desorption, the couple E$_{bind}$ and $\nu$, are shown in table \ref{<Edes>} and we provide a comparison between the experimental fit obtained with the literature values. 

\subsection{Deriving kinetic parameters for desorption}
\label{kinetic_param}

The kinetics of the desorption is described fitting the rate of de\-sorp\-tion R(t) [molecules cm$^{-2}$ s$^{-1}$], which is the actual measured quantity, against the Wigner-Polanyi equation \citep{RedheadVacuum62, CarterVacuum62}:

\begin{equation}
R(t)=-\frac{dN}{dt}= k(T)\cdot N^i , 
\label{WPt}
\end{equation}  

\noindent where \textit{N} is the number of adsorbed molecules on the surface [molecule cm$^{-2}$] and \textit{i} is the order of desorption. The order of desorption is 0 for ice mantle desorption, 1 for surface desorption and 2 for reactive desorption \citep{RedheadVacuum62, CarterVacuum62}. Contrary to certain authors \citep{BolinaSurSc05, BolinaJCP05}, we do not consider non-integer orders of desorption. 
In the case of zero-order desorption, the rate constant \textit{k(T)} is expressed in units of [monolayer.cm$^{-2}$.s$^{-1}$] , indicating a constant desorption rate independent of the surface coverage. The desorption rate \textit{k(T)} is described by an Arrhenius law since desorption is an activated process:

\begin{equation}
k(T)=\nu \cdot exp\left(-\frac{E_{des}}{RT}\right),
\label{Arh}
\end{equation}

\noindent where $E_{des}$ is the activation energy for desorption [J K$^{-1}$ mol$^{-1}$] and $\nu$ [s$^{-1}$] the pre-exponential factor. As a first approximation, we describe desorption as an elementary process and $\nu$ is considered temperature-independent. The likely series of activation processes (rotation diffusion, breaking of intermediate bonds, ...) hidden in this pre-exponential factor could introduce a temperature dependence. Determining this pre-exponential factor is as important as determining the desorption energy \citep{DoroninJCP15, LunaVacuum15}. Most of the time the desorption parameters are not derived from isothermal experiments but rather from TPD experiments, where the temperature ramp is increased linearly T~=~T$_0$~+~$\beta$t, and $\beta$ is the temperature heating rate. Although, this formally just implies a change of variable in Eq.~\ref{WPt}:

\begin{equation}
R(T)=-\frac{dN}{dT}= \frac{k(T)}{\beta}\cdot N^i, 
\label{WPT}
\end{equation}

\noindent the coupling of time and temperature becomes complex when other processes that also depend on these factors, such as ice crystallisation, occur simultaneously. 

We note that the TPD technique is used to derive the desorption energy for a given system, rather than the binding energy. While the binding energy refers to the energy required to maintain a bond between a molecule and a surface, the desorption energy specifically refers to the energy needed to break this bond and release the molecule. Although these energies may be similar or even identical in certain systems, they remain conceptually distinct. In this paper, we assume \(E_{bind} = - E_{des}\), and we report values derived from Eq. \ref{WPT} as binding energies.

To obtain the binding energies and pre-exponential factors from the TPD curves from Equation~\ref{WPT}, a standard minimisation of $\chi^2$ is used, which reflects the sum of the squares of the differences between the experimental and the calculated profiles. The values of the desorption rate as a function of the temperature and the surface coverage were obtained directly from the experimental data. In order to work with only the binding energy as a free parameter during the fit, we fixed the pre-exponential factor constant. Optimisation of both parameters was then performed to reduce the $\chi^2$ value. We sourced the pre-exponential factor using the recommended values provided in Table 2 of \cite{Minissale_review_22} which were derived for monolayer desorption on compact water ices. These pre-exponential factor values were determined with Redhead Transition State Theory (Redhead-TST) which offer enhanced accuracy over those estimated through the Hasagawa equation. It is important to note, however, that the pre-exponential factors derived by \cite{Minissale_review_22} are determined for monolayer coverages (n = 1), whereas our study is focused on determining the binding energies for multilayer coverages (n = 0). This discrepancy in coverage models may account for the observed deviations between our results and the literature values reported in Table \ref{<Edes>}.

\subsection{Pure ice desorption}

\subsubsection{CO-like molecule desorption: Ar, CO}

\begin{figure}[h]
   \centering
   
   \includegraphics[width = 9 cm]{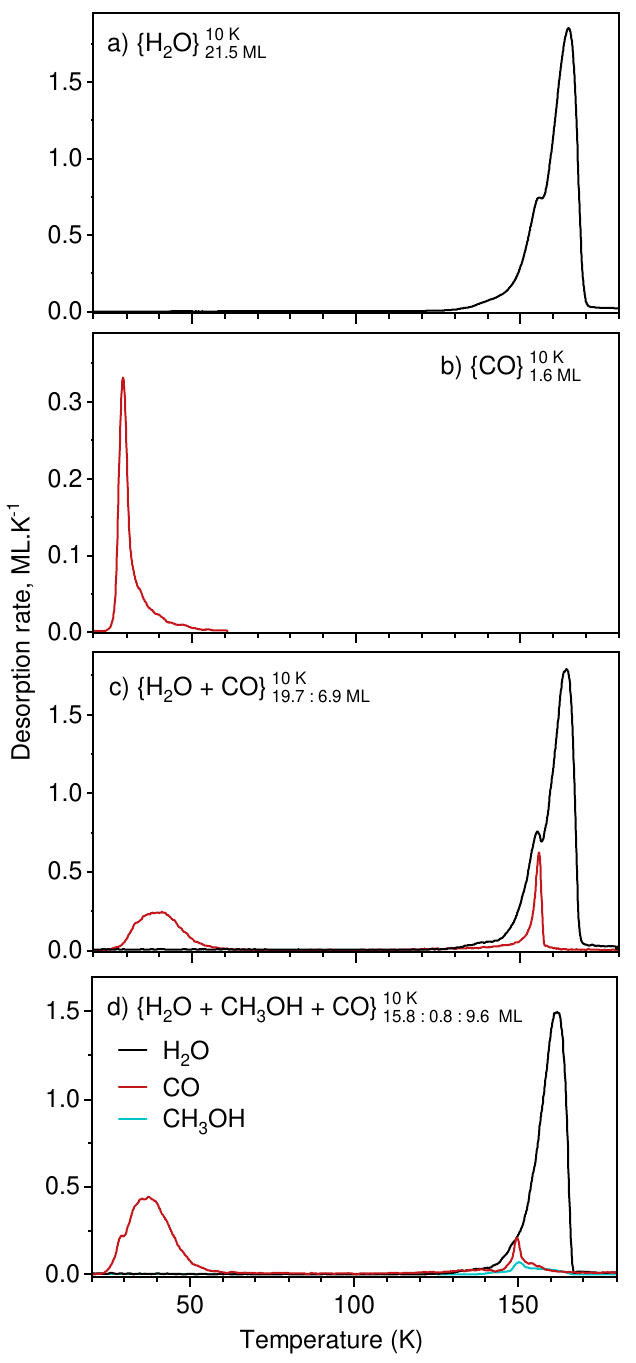}
   \caption{
   Desorption curves for experiments involving CO and H$_2$O ice mixtures: (a) Pure H$_2$O and (b) CO ices. (c) A binary mixture of H$_2$O and CO. (d) Ternary mixture of H$_2$O, CH$_3$OH, and CO. The heating rate is 0.2~K.s$^{-1}$. The deposition temperature and the deposition dose for each component in the mixture are listed for each experiment.
   }
   \label{Fig_H2O_CO}
    \end{figure}

\begin{figure}[ht]
   \centering
   \includegraphics[width = 9 cm]{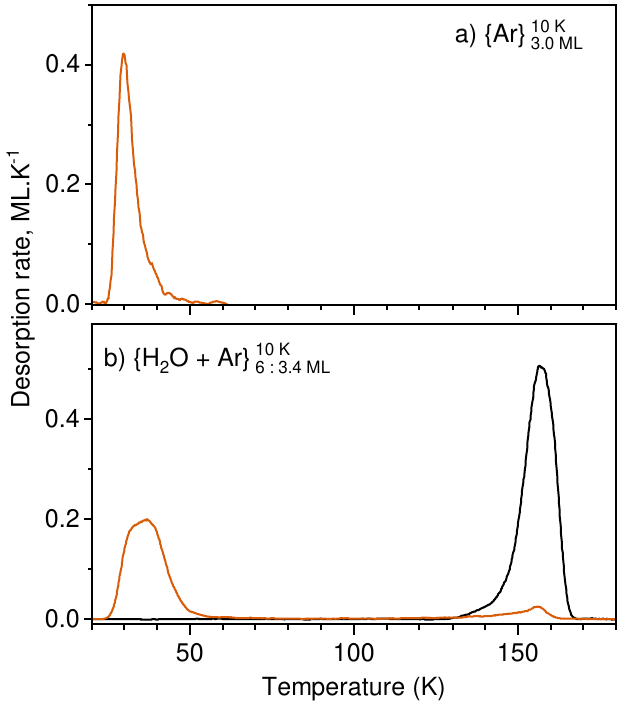}
   \caption{
   Desorption curves for TPD experiments involving Ar and H$_2$O ice mixtures: (a) Pure Ar and (b) a binary mixture of H$_2$O and Ar. The heating rate is 0.2~K.s$^{-1}$. The deposition temperature and the deposition dose for each component in the mixture is listed in for each experiment.
   }
   \label{Fig_H2O_Ar}
    \end{figure}

Both Argon and CO exhibit similar binding energy values and desorb within a temperature range of 20 - 60 K, as shown in Fig.\ref{Fig_H2O_Ar}a) for Argon and Fig.\ref{Fig_H2O_CO}b) for CO. To model this, we fitted the exponential behaviour of the TPD  curves — using a temperature cutoff of 27 K for Argon and 28 K for CO — to the zeroth-order Polanyi-Wigner equation. This yielded E$_{bind}$ values of 933 $\pm$ 96 K and $\nu_0$ of 3.8 $\pm$ 3.0 $\times$ 10$^{13}$ ML.s$^{-1}$ for Argon, and E$_{bind}$ values of 1040 $\pm$ 104 K and $\nu_0$ of  8.7 $\pm$ 2.0 $\times$ 10$^{14}$ ML.s$^{-1}$ for CO. The experimental results and literature values are in good agreement for Argon. For CO, the binding energy is slightly lower than the literature value reported by \cite{Minissale_review_22}, which can be an effect of the multilayer surface coverage.

\subsubsection{Intermediate species desorption: CO$_2$}

CO$_2$ desorbs within the temperature range of 70 - 90 K, as shown in Fig.~\ref{Fig_H2O_CO2}b). We applied a zero-order desorption fit in the exponential regime of the data (T $<$ 77 K) yielding to E$_{bind}$ = 2750 $\pm$ 253 K and $\nu$ = 2.2 $\pm$ 2.4 $\times$ 10$^{16}$ ML.s$^{-1}$. We fixed the pre-exponential factor based on the values recommended in table 2 of \cite{Minissale_review_22}. However, the best fit was found to be much lower than those values for both parameters. The best fit in our experimental data are in agreement with values reported by \cite{Ulbricht06}, E$_{bind}$ = 2766 K and $\nu$ = 6 $\times$ 10$^{14}$ ML.s$^{-1}$.

\subsubsection{H$_2$O-like molecule desorption: NH$_3$, CH$_3$OH, H$_2$O}

The desorption curve of water ice is presented in Fig. \ref{Fig_H2O_CO}a), showing a desorption temperature range of 130 - 180 K. The salient feature observed at approximately 154~K marks the irreversible phase transition from amorphous solid water to crystalline ice. This transition is crucial as it establishes the location of the first desorption peak related to trapped species within the water ice, commonly referred to as the "volcano peak".

For the curve fitting, two separate temperature regimes were considered to obtain desorption parameters. For amorphous solid water, the temperature was constrained to T $<$ 153 K, yielding E$_{bind}$ = 6638 $\pm$ 110 K and $\nu$ = 8.7 $\pm$ 1.1 $\times$ 10$^{17}$ ML.s$^{-1}$. The parameters for crystalline ice were determined within a narrower temperature range of 158 $<$ T $<$ 162 K, which increases the uncertainties of the obtained values. For crystalline ice, we found E$_{bind}$ = 6829 $\pm$ 137 K and $\nu$ = 9.0 $\pm$ 1.1 $\times$ 10$^{14}$ ML.s$^{-1}$. The pre-exponential factors were set to the values reported by \cite{smith_JPCA_2011}, which provide data for both amorphous and crystalline water ice desorption.

The NH$_3$ desorption curve, displayed in Fig. \ref{Fig_H2O_NH3}a), reveals a desorption peak in the temperature range of 70 - 100 K. The curve was fitted using the Polanyi-Wigner equation in a regime well-described by exponential behaviour, with a temperature cut-off set at T $<$ 87 K. The derived best-fit parameters are E$_{bind}$ = 3065 $\pm$ 251 K and $\nu$ = 1.0 $\pm$ 1.3 $\times$ 10$^{13}$ ML.s$^{-1}$. We initially constrained the pre-exponential factor based on the values recommended in Table 2 of \cite{Minissale_review_22}. Notably, our best-fit values for both E$_{bind}$ and $\nu$ were substantially lower than these recommended values. This discrepancy was also observed in the curve fitting for CO$_2$. Despite the lower fitting parameters, our results are in good agreement with the multilayer desorption values reported by \cite{Ulbricht06}, specifically E$_{bind}$ = 3007 K and $\nu$ = 5 $\times$ 10$^{13}$ ML.s$^{-1}$.

In Fig. \ref{Fig_H2O_CH3OH}a, we present the TPD curve for pure CH$_3$OH ice, exhibiting desorption between 120 - 150~K. Our zeroth-order kinetics fitting using the Polanyi-Wigner equation yielded E$_{bind}$ = 5531 $\pm$ 113 K and $\nu$ = 8.8 $\pm$ 1.9 $\times$ 10$^{16}$ ML.s$^{-1}$. 

\subsubsection{Semi-volatile molecule desorption: NH$_4^+$HCOO$^-$}

Figure \ref{Fig_H2O_NH4HCOO}(a) displays the TPD curve for the desorption of ammonium formate, NH$_4^+$HCOO$^-$. As a semi-refractory molecule, NH$_4^+$HCOO$^-$ desorbs at higher temperatures than water ices, in the range of 180 - 240 K. Using zeroth-order kinetics for the exponential part of the curve, specifically at T $<$ 200 K, we obtained the best-fit values as E$_{bind}$ = 9151 $\pm$ 197 K and $\nu$ = 1.6 $\pm$ 5.9 $\times$ 10$^{15}$ ML.s$^{-1}$.

\begin{table*}[ht]
\caption{Comparison of kinetic parameters for desorption of multi-layer coverages of various species from this work and the literature.}
\label{<Edes>}
\centering
\begin{threeparttable}
\begin{tabular}{lccccc}
\toprule
 & \multicolumn{2}{c}{This work} & \multicolumn{2}{c}{Literature values} &  \\ 
\cmidrule(lr){2-3} \cmidrule(lr){4-5}
\multirow{1}{*}{Species} & E$_{bind}$ (K) & $\nu$ (ML.s$^{-1}$) & E$_{bind}$ (K) & $\nu$ (s$^{-1}$) & \multirow{1}{*}{Reference} \\
\midrule
Ar & 933 $\pm$ 96 & 3.8 $\pm$ 3.0 $\times$ 10$^{13}$ & 864 & 1.84 $\times$ 10$^{13}$ & \cite{Minissale_review_22}  \\
CO & 1040 $\pm$ 104 & 8.7 $\pm$ 2.0 $\times$ 10$^{14}$ & 1390 & 9.14 $\times$ 10$^{14}$ & \cite{Minissale_review_22}  \\
CO$_2$ & 2750 $\pm$ 253 & 2.2 $\pm$ 2.4 $\times$ 10$^{14}$ & 3196 & 6.81 $\times$ 10$^{16}$ & \cite{Minissale_review_22} \\
&   &  & 2766 & 6 $\times$ 10$^{14}$ & \cite{Ulbricht06} \\
NH$_3$ & 3065 $\pm$ 251 & 1.0 $\pm$ 1.3 $\times$ 10$^{14}$ & 2760 & 1.94 $\times$ 10$^{15}$  & \cite{Minissale_review_22}  \\
  &     &   & 3007 & 5 $\times$ 10$^{13}$  & \cite{Ulbricht06}\\
CH$_3$OH & 5531 $\pm$ 113 & 8.8 $\pm$ 1.9 $\times$ 10$^{16}$ & 5512 & 3.18 $\times$ 10$^{17}$ & \cite{Minissale_review_22}  \\
H$_2$O$_{am}$ & 6638 $\pm$ 134 & 8.7 $\pm$ 1.1 $\times$ 10$^{17}$ & 6560 & 1.1 $\times$ 10$^{18}$ & \cite{smith_JPCA_2011} \\
H$_2$O$_{cryst}$ & 6829 $\pm$ 137 & 9.0 $\pm$ 1.1 $\times$ 10$^{17}$ & 6722 & 1.3 $\times$ 10$^{18}$ & \cite{smith_JPCA_2011} \\
NH$_4$$^+$HCOO$^-$ & 9151 $\pm$ 197 & 1.6 $\pm$ 5.9 $\times$ 10$^{18}$ & 9450 & 2.0 $\times$ 10$^{18}$ & \cite{Ligterink2023} \\
\bottomrule
\end{tabular}
\end{threeparttable}
\end{table*}

\subsection{Binary ice mixtures}

\subsubsection{CO-like molecule desorption: Ar, CO}

In Fig. \ref{Fig_H2O_CO}(c), the TPD spectrum from an experiment on a binary mixture of H$_2$O and CO is shown (refer to Exp. 18 in Table \ref{table_overview}). This spectrum clearly delineates three primary desorption features: two associated with CO (in red) and a singular peak from H$_2$O (in black). The CO desorption occurring between 20 to 60~K is identified as the surface peak, indicative of CO molecules interacting with both the external water ice surface and the initial few layers of the ice bulk. Conversely, the desorption of CO observed between 140 to 160~K, labelled the volcano peak, corresponds to the release of molecules initially trapped within the water ice bulk. This release is driven by the phase transition of water ice from amorphous to crystalline. Structural changes, especially the formation of cracks during this transition \citep{MayI_JCPA_2013}, facilitate the liberation of these entrapped molecules.

Panel b) of Fig.\ref{Fig_H2O_Ar} presents the desorption characteristics of a binary mixture of H$_2$O and Ar (refer to Exp. 24). Given the volatility of Argon and binding energy, akin to CO (as per Table\ref{<Edes>}), it is anticipated to exhibit similar desorption trends in a binary mixture. The TPD spectrum delineates two distinct Argon desorption peaks: a low-temperature surface peak and a broad volcano peak at the higher temperature regime. However, we observe a notable difference in the crystallisation of water between H$_2$O~:CO and H$_2$O:Ar experiments. Although the TPD curve for water in Fig.\ref{Fig_H2O_CO} exhibits a well-defined crystallisation curve highlight by the "bump" around 154 K, this characteristic is absent from the TPD curve for water in the binary mixture with Argon. This suggests that Argon influences the crystallisation kinetics of water. Moreover, we find the peak of trapped Argon to be larger and more comparable to co-desorption with water than a volcano peak related to the water crystallisation.

    \begin{figure}[ht]
   \centering
   \includegraphics[width = 9 cm]{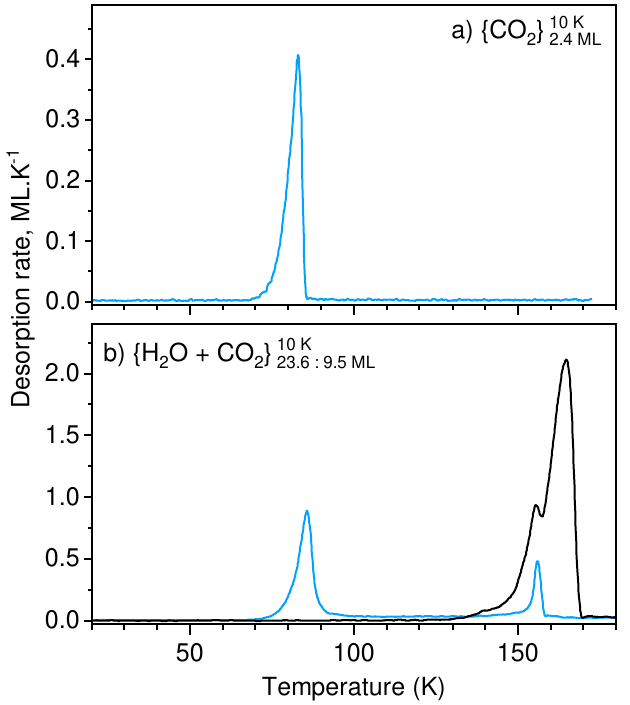}
   \caption{
   Desorption curves for TPD experiments involving CO$_2$ and H$_2$O ice mixtures: (a) Pure CO$_2$ and b) a binary mixture of H$_2$O and CO$_2$. The heating rate is 0.2 K.s$^{-1}$. The deposition temperature and the deposition dose for each component in the mixture are listed in each experiment.}
   \label{Fig_H2O_CO2}
    \end{figure}

\subsubsection{Intermediate species desorption: CO$_2$}

Figure \ref{Fig_H2O_CO2} presents the desorption curve for a H$_2$O and CO$_2$ mixture (refer to Exp.26). CO$_2$ exhibits two desorption peaks. The surface peak, between 70 and 100~K, corresponds to the desorption from molecules interacting with the surface and the initial layers of compact water ice. The subsequent peak at 156~K constitutes the volcano peak, occurring during the crystallisation of water ice. Additionally, we observe some ice loss between these two primary peaks in the CO$_2$ experiments, a characteristic not seen in experiments with other species like CO or Argon.

\subsubsection{H$_2$O-like molecule desorption: NH$_3$, CH$_3$OH}

\begin{figure}[ht]
   \centering
   \includegraphics[width = 9 cm]{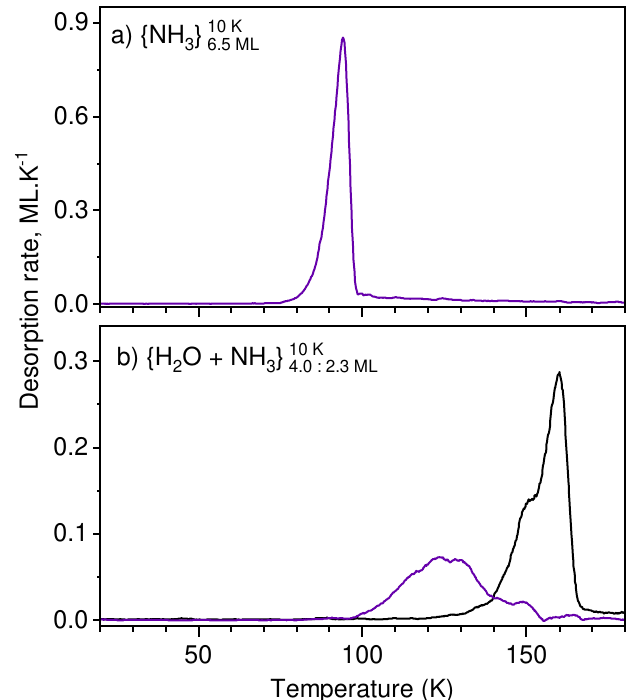}
   \caption{
   Desorption curves for TPD experiments involving NH$_3$ and H$_2$O ice mixtures: (a) Pure NH$_3$ and (b) a binary mixture of H$_2$O and NH$_3$. The heating rate is 0.2 K.s$^{-1}$. The deposition temperature and the deposition dose for each component in the mixture are listed in each experiment.
   }
   \label{Fig_H2O_NH3}
    \end{figure}

Figure \ref{Fig_H2O_NH3} displays the TPD spectra of pure NH$_3$ and a H$_2$O : NH$_3$ mixture (Exps. 5 and 32). The TPD curve of pure NH$_3$ reveals a desorption peak at 90~K. In contrast, the TPD curve for ammonia in the water mixture indicates a broader desorption range from 100 to 155~K for NH$_3$. The formation of hydrogen bonds between NH$_3$ and H$_2$O, and their incorporation into its structure, results in a significant shift in binding energy values compared to multilayer coverages. The binding energy of NH$_3$ on diverse water ice surfaces is a key parameter for astrophysical models and it presents a distribution of values depending on the substrate \citep{Suresh2023, Tinacci2022}. 

\begin{figure}[ht]
   \centering
   \includegraphics[width = 9 cm]{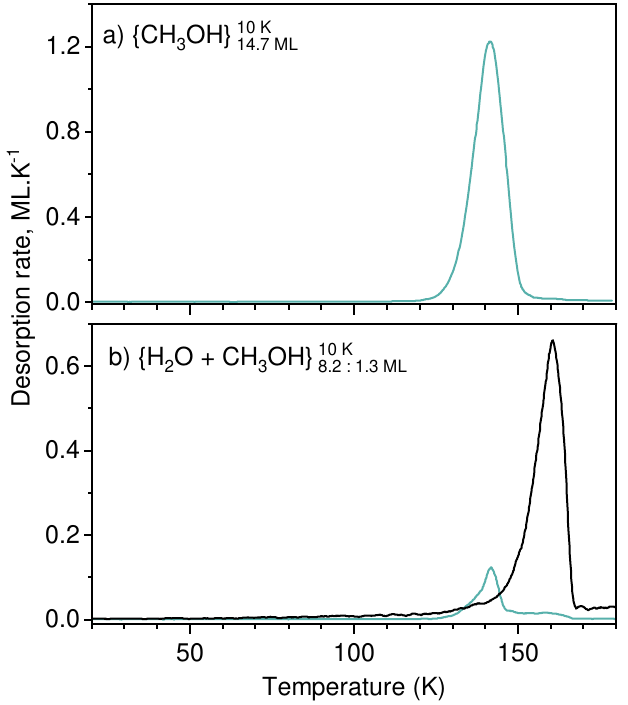}

   \caption{
   Desorption curves for TPD experiments involving CH$_3$OH and H$_2$O ice mixtures: (a) Pure CH$_3$OH and (b) a binary mixture of H$_2$O and CH$_3$OH. The heating rate is 0.2 K.s$^{-1}$. The deposition temperature and the deposition dose for each component in the mixture are listed in each experiment.
   }
   \label{Fig_H2O_CH3OH}
    \end{figure}

The TPD curve for a H$_2$O:CH$_3$OH binary mixture is shown in Fig. \ref{Fig_H2O_CH3OH} (refer to Exp. 27). Panel a) displays the pure CH$_3$OH desorbing around 140~K. In contrast, panel b) exhibits a methanol desorption component in the binary mixture within the same temperature range as pure methanol, centred at 140~K, indicative of the surface peak. A secondary desorption component, extending until 165~K, is related to the co-desorption of methanol and water. We also point out that the presence of methanol affects the water crystallisation kinetics, as the water desorption curve lacks the characteristic crystallisation "bump" for the phase transition.

\subsubsection{Semi-volatile molecule desorption: NH$_4^+$HCOO$^-$}

\begin{figure}[ht]
   \centering
   \includegraphics[width = 9 cm]{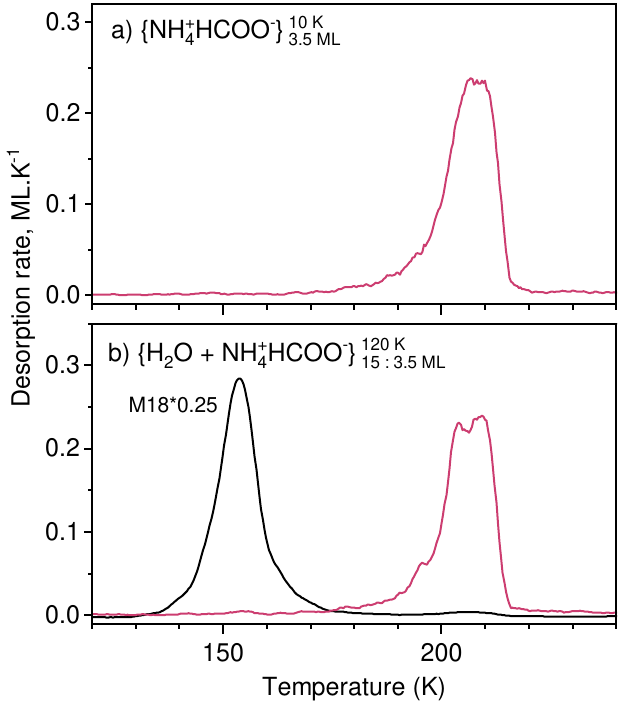}
   \caption{
   Desorption curves for TPD experiments involving NH$_4$$^+$HCOO$^-$ and H$_2$O ice: (a) Pure NH$_4$$^+$HCOO$^-$ and (b) a film of 15 ML of H$_2$O deposited bellow 3.5 ML of NH$_4$$^+$HCOO$^-$. The heating rate is 0.2 K.s$^{-1}$. In panel (b), the water desorption peak is scaled to align with the  proportion of the salt peak. The deposition temperature and the deposition dose for each component in the mixture are listed in each experiment. 
   }
   \label{Fig_H2O_NH4HCOO}
    \end{figure}

Figure \ref{Fig_H2O_NH4HCOO} presents the TPD spectra for the ammonium formate salt, NH$_4^+$HCOO$^-$, deposited without water ice (panel a) and with 15 ML of water ice deposited before the salt and serving as a substrate (panel b). 

In the experiment showed in panel (b), the underlying water ice appears to exert no significant influence on the desorption behaviour of the ammonium formate salt. The salt desorbs at a temperature consistent with the pure salt, with no shift in the peak temperature due to the presence of the water ice substrate. 

Moreover, we point out that in this particular experiment, we deposited the amorphous water ice via background deposition before depositing the salt. This factor accounts for the difference in the water TPD curve in Fig. \ref{Fig_H2O_NH4HCOO} panel (b), as compared to other experiments involving amorphous compact water ice.

\subsection{Ternary mixtures: H$_2$O + CH$_3$OH + CO}

Panel (d) of Figure \ref{Fig_H2O_CO}  presents the TPD spectrum for the desorption of the ternary mixture H$_2$O:CH$_3$OH:CO (refer to Exp. 35). Two distinctive peaks for CO are observed, the surface peak at a lower temperature range, and one in the higher temperature range, which are the same features identified in binary mixtures. However, in the binary mixture, the volcano peak reaches its peak around 154 K, while in the ternary mixture, the peak is broader and occurs around 145 K. As demonstrated in the binary mixture of H$_2$O:CH$_3$OH, the inclusion of methanol influences the crystallisation kinetics of water ice and it is also affecting the shape of the volcano peak. The absence of the "bump" in the water desorption curve of the ternary mixture further supports this observation. Analogous behaviours have been reported for ternary mixtures of water, methanol, and either OCS or CO$_2$ \citep{BurkeBrown15}.

\section{Results: Desorption yields}
\label{results_yields}

The thermal desorption processes of interstellar ice analogues are comprehensively analysed by examining both the kinetics, detailing the rate and temperature-dependent characteristics of desorption, and the desorption yields, quantifying the amount of material released. This section focuses on the desorption yields for ice mixtures of compact amorphous water ice, which relate to the number of molecules released during distinct desorption peaks. These yields are obtained from the integrated areas of the specific TPD curve features: the surface peak, the volcano peak, and a co-desorption peak with H$_2$O, if present. 

In this study, the trapped fraction, $x_{\text{trapp}}$, is defined as the amount of the species (\textit{x}) within water ice (released in both the volcano and co-desorption when water desorbs) over the total amount of \textit{x} deposited:

\begin{equation}
    x_{\text{trapp}} = \frac{x_{volc} + x_{codes}}{x_{tot}}.
    \label{trapp_eq}
\end{equation}

The trapping efficiency of water, $x_{\text{eff-water}}$, is defined as the amount of the species, present in both the volcano and co-desorption peaks, divided by the total amount of H$_2$O deposited:

\begin{equation}
    x_{\text{eff-water}} = \frac{x_{volc} + x_{codes}}{H_2O_{tot}}.
    \label{x_eff}
\end{equation}

We also point out that the co-desorption peak for high volatile species is predominant in ices with thicknesses exceeding 100~ML \citep{MayI_JCPA_2013}. For binary mixtures of H$_2$O:CO and CO$_2$ ranging from 3~to 80~ML in our experiments, no co-desorption peak was detected. Thus, we can simplify equations \ref{trapp_eq} and \ref{x_eff}, as their $x_{codes}$ equals zero: 

\begin{equation}
    CO_{\text{trapp}} = \frac{CO_{volc}}{CO_{tot}}
    \label{CO_trapp}
,\end{equation}

\begin{equation}
    CO_{\text{eff-water}} = \frac{CO_{volc}}{H_2O_{tot}}
    \label{CO_eff}
.\end{equation}

Lastly, both CH$_3$OH and NH$_3$ exhibit multilayer desorption features that blend closely with the volcano peak, making them difficult to separate. To accurately measure the desorption yields, we focused exclusively on the co-desorption peak that occurs during the desorption of crystalline water. Our approach involved initially fitting both the individual desorption peak of methanol or ammonia and their co-desorption peak with water. Following this, we integrated the peaks and determine their respective desorption yields considering $x_{volc}$ equal zero.

In order to discuss the results, we plotted the trapped fraction, \textit{x}$_{\text{trapp}}$, and the trapping efficiency of water, \textit{x}$_{\text{eff-water}}$, as a function of the H$_2$O ice coverage of the experiments (Fig.~\ref{Fractions}). The results for each desorption category are discussed in the sections below.

\subsection{CO-like molecule desorption: CO, Ar}

\begin{figure*}[h]
   \centering
   \includegraphics[width = 18 cm]{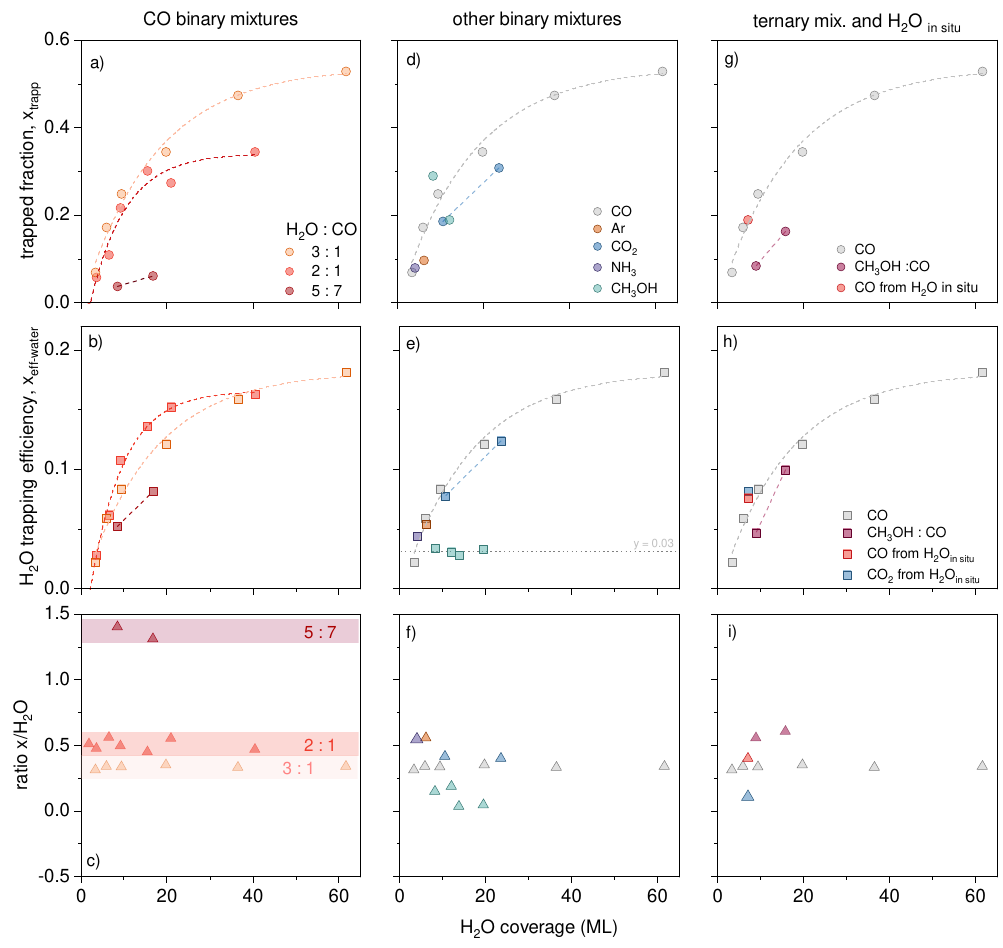}
   \caption{
   Comparative analysis of desorption yields in various ice mixtures. The trapped fraction (a, d, g) and the H$_2$O trapping efficiency (b, e, h) as functions of H$_2$O ice coverage, across CO binary mixtures, other binary mixtures, and for ternary mixtures and the \textit{in situ} production of H$_2$O, respectively. Panels (c), (f), and (i) show the deposition ratio for the corresponding systems. The dashed lines represent an exponential fit to the data, intended solely to assist visual interpretation. Dashed lines represent exponential and linear fits to the data, provided to assist visual interpretation and it is not meant to imply an underlying model.}
   \label{Fractions}
    \end{figure*}

Figure~\ref{Fractions} displays the desorption yields obtained from a series of experiments varying H$_2$O~:~CO ratios (refer to Exp.8-22 of Table \ref{table_overview}). In Panel a), the trapped fraction is shown, while Panel b) shows the trapping efficiency of water. Panel c) presents the ratio of deposited CO to H$_2$O, with specific average ratios of each experiment for H$_2$O:CO being 2 : 1, 3 : 1, and 5 : 7. Distinct trends in the trapped fraction are observed for H$_2$O : CO ratios of 2 : 1, 3 : 1, and 5:7, as indicated in Panel a). With a H$_2$O:CO ratio of 2:1 (denoted by red points), the trapped fraction demonstrates an increase with deposition dose, stabilising near a value of $CO_{\text{trapp}}\approx$ 0.5. Based on Eq. \ref{CO_trapp}, the trapped fraction for CO is defined as the amount of CO desorbing with H$_2$O in the volcano peak over the total CO deposited. Consequently, a $CO_{\text{trapp}}\approx$ 0.5 suggests a balanced desorption of molecules from both the surface and volcano peaks. For a H$_2$O:CO ratio of 2:~3 (dark red points), the lower trapped fraction can be attributed to a predominance of CO molecules desorbing at the surface peak relative to the volcano peak.

In Fig.\ref{Fractions}(b), the trapping efficiency of water for CO, as defined by Eq.\ref{CO_eff}, is plotted. This graph indicates a clear trend: as the H$_2$O ice coverage increases, so does the amount of CO desorbing at the volcano peak, regardless of the H$_2$O:CO ratios studied. While there are some differences across the data, a steady increase with water ice coverage is consistently observed. The H$_2$O:CO ratios of 2:1 and 3:1 appear to reach a stable efficiency value, with CO$_{\text{eff-water}}$ nearing 0.2. The trapping efficiency indicates the proportion of CO released at the peak compared to the initially deposited H$_2$O volume. The trend suggests that a H$_2$O ice layer may trap a maximum of about 20\% of CO based on its thickness. For example, an ice layer of 60 ML of H$_2$O can trap a maximum of 12 ML of CO (60 ML $\times$ 0.2 = 12 ML). Furthermore, when the H$_2$O:CO ratio is above 0.2, as in experiments with a higher CO content than water (H$_2$O:CO = 5:7), the excess CO molecules desorb at the peak.

Panel d) of Fig.\ref{Fractions} exhibits the trapped fraction for H$_2$O:~Ar (in orange). We performed this experiment in a low thickness regime, where the deposited water ice has a coverage of of less than 30~ML of H$_2$O. Although we cannot conclusively state that the plateau behaviour of H$_2$O:Ar binary mixtures would follow the same trend observed for H$_2$O:~CO thick ices (> 30 ML of H$_2$O), in the low thickness regime (< 30 ML H$_2$O), compact amorphous water ice traps a comparable amount of CO and Ar.

\subsection{Intermediate species desorption: CO$_2$}

Figure \ref{Fractions} panels d) and e) exhibit the desorption yields for H$_2$O:~CO$_2$ binary ice mixtures, in blue. We conducted two experiments within a low-thickness regime, with less than 30 ML of H$_2$O ice deposited. 
In this regime, CO$_2$ entrapment aligns with the same trend observed in the CO binary mixture experiments (highlighted by light gray points for comparison). Both the trapped fraction and the trapping efficiency of water increase with deposition dose, possibly approaching plateau values similar to those observed with CO: $CO_2$$_{\text{trapp}}$ $\approx$ 0.5 and $CO_2$$_{\text{eff-water}}$ $\approx$ 0.2.

\subsection{H$_2$O-like molecule desorption: CH$_3$OH, NH$_3$}
\label{des_yield_metOH}

In the H$_2$O:CH$_3$OH binary mixture experiments, we expanded the range of variables compared to other mixtures by varying the deposition dose, layered and homogeneous ice mixtures, the H$_2$O:CH$_3$OH ratio, and the deposition temperature (both 10 and 90 K). A comprehensive overview of these experiments can be found in Table \ref{table_overview} (Exp. 27 - 30).

Panel (d) of Figure \ref{Fractions} presents the trapped fraction for methanol in these mixtures (green dots). There is considerable variation among these values, with some data points even exceeding the displayed range. However, the trapping efficiency of water for methanol showed in panel e) consistently falls within the $CH_3OH_{\text{eff-water}}\approx$ 0.03 range across all data, despite differences in ice morphology due to varying deposition parameters.

In the case of the H$_2$O~:~NH$_3$ binary mixtures, only one experiment was conducted, yielding a water trapping efficiency for ammonia of $NH_3$$_{\text{eff-water}}=$ 0.044. Given that CH$_3$OH and NH$_3$ belong to the same desorption category, we can anticipate a comparable behaviour from NH$_3$ to that of methanol. We therefore predict that despite different water-ice structures, the trapping efficiency of water for ammonia is expected to have values around 0.05. 

\subsection{Semi-volatile molecule desorption: NH$_4^+$HCOO$^-$ }

In this study, we adopt the definition of `trapping' as described by \cite{Collings_2004MNRAS_survey}, which we understand as the process where a species is physically incorporated into the water ice matrix. This incorporation is significant because it means the interaction between the adsorbed species and the water ice surface is not the primary factor influencing the temperature at which the species desorbs. For molecules with binding energies lower than that of water ice (e.g. CO-like molecules, intermediate species), it is straightforward to deduce the amount of molecules physically trapped within the ice by measuring the observed volcano and co-desorption peaks in the TPD curves. For semi-volatile species, however, adhering strictly to this definition does not conclusively indicate the absence of trapping. These species desorb at temperatures higher than water ice, making difficult the determination of whether they were similarly trapped within the ice at the moment of water ice desorption.

Nevertheless, it is essential to avoid over-complicating the analysis and the key takeaway from our experimental results is that NH$_4^+$HCOO$^-$ exhibited no volcano peak or co-desorption with water, thus the desorption yields are zero and all deposited ammonium formate is desorbing at the same temperature as the pure salt, as shown in Fig.\ref{Fig_H2O_NH4HCOO}. 

\subsection{Ternary mixtures}

Panels (g) and (h) of Figure \ref{Fractions} compare the quantities of CO trapped in H$_2$O:CH$_3$OH:CO ternary mixtures (purple dots) with those found in H$_2$O : CO = 3:~1 binary mixtures. Although the inclusion of methanol in the ternary mixture does impact the kinetics of the crystallisation process during the phase transition of water ice (as demonstrated in Fig. \ref{Fig_H2O_CO}c), it does not significantly alter the amount of CO entrapped within water. This is reflected in the similar range of trapping efficiency values of water for CO in both ternary and binary mixtures.

\subsection{H$_2$O formed \textit{in situ}} 

Experiments were conducted to ascertain if a highly compact water film, formed \textit{in situ}, exhibited trapping properties comparable to those of predeposited water. In these experiments, water was synthesised through the co-deposition of molecular oxygen and atomic hydrogen. The atomic hydrogen was generated by the dissociation of H$_2$ molecules, facilitated by a microwave discharge installed in the right beam of the VENUS setup. This process induced the following reactions on the cold surface:

\begin{equation}
\text{O}_2 + \text{H} \rightarrow \text{O}_2\text{H},
\end{equation}
\begin{equation}
\text{O}_2\text{H} + \text{H} \rightarrow \text{H}_2\text{O}_2,
\end{equation}
\begin{equation}
\text{H}_2\text{O}_2 + \text{H} \rightarrow \text{H}_2\text{O} + \text{OH},
\end{equation}
and
\begin{equation}
\text{OH} + \text{H} \rightarrow \text{H}_2\text{O}.
\end{equation}


This procedure to form H$_2$O presents a number of challenges, primarily due to the fact that in addition to H$_2$O, H$_2$O$_2$ is produced, and the ratio of H$_2$O~/~H$_2$O$_2$ is dependent on the initial flux of hydrogen atoms.
In a more reducing environment, H$_2$O$_2$ will convert to H$_2$O, as shown above.
In an oxidising environment, where there are a high concentration of oxygen and hydroxyl (OH) species, there is a greater propensity for the formation of oxygenated water, H$_2$O$_2$.
In the experiment that we conducted, we successfully achieved a ratio of H$_2$O to H$_2$O$_2$ of 4 : 1.
Moreover, we co-deposited $^{13}$CO into the mixture as ${O_2 + H + ^{13}CO}^{10K}$ to test the trapping properties of the  H$_2$O being formed \textit{in situ}. Thus, even though we are using the isotope $^{13}$CO to more clearly distinguish the carbonated molecules, a multitude of secondary species emerge during the experiment. It is beyond the scope of this work to discuss their formation routes. The discussion that follows focuses on the trapping properties of a film of H$_2$O that was synthesised \textit{in situ}. 

\begin{figure}[ht]
   \centering
   \includegraphics[width = 9 cm]{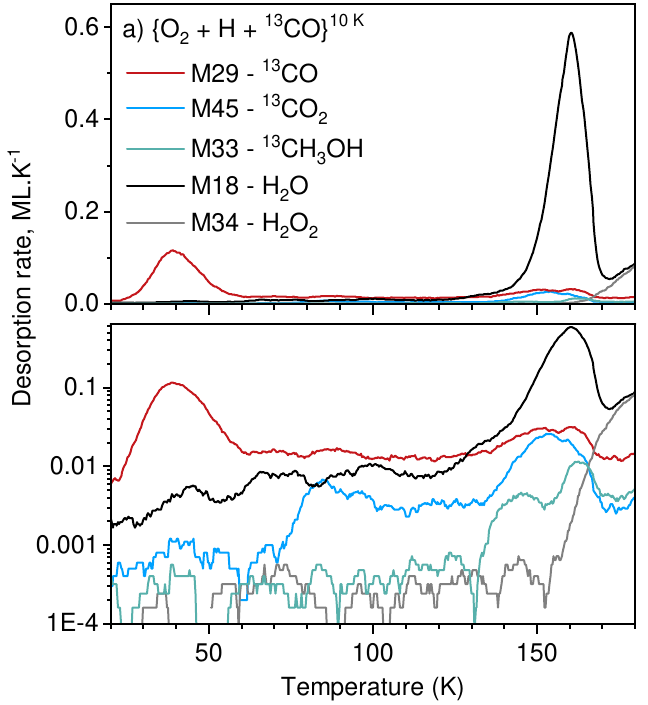}
   \caption{
   Desorption curves for TPD experiments involving the formation of H$_2$O {in situ} through the co-deposition of $\{O_2 + H + ^{13}CO\}$. Lower panel is the log-scale of panel (a).}
   \label{Fig_H2Oinsitu_CH3OH_X}
    \end{figure}

Figure \ref{Fig_H2Oinsitu_CH3OH_X} presents the results of this experiment (refer to Exp. 34 of Table \ref{table_overview}). We trace the desorption curves of the following fragments: m/z = 29 for $^{13}$CO, m/z = 45 for $^{13}$CO$_2$, m/z = 33 for $^{13}$CH$_3$OH,  m/z = 18 for H$_2$O and m/z = 34 for H$_2$O$_2$. In the lower temperature range, the desorption profile of $^{13}$CO aligns with the profile observed of the surface peak of CO in water and CO binary mixture studies (see Fig. \ref{Fig_H2O_CO} for a direct comparison). Desorption of $^{13}$CO$_2$ begins around 70 K, emerging as a product from reactions involving $^{13}$CO and OH radicals as reported by \citep{Ioppolo2011}. The high-temperature range between 130 and 170 K shows the desorption of H$_2$O molecules, as well as $^{13}$CO, $^{13}$CO$_2$, and $^{13}$CH$_3$OH trapped within the water ice structure. Following the peak of water desorption, we also see the beginning of H$_2$O$_2$ sublimation.

In panels (g) and (h) of Fig.\ref{Fractions}, we present the trapped fraction and the trapping efficiency of water for the ${O_2 + H + ^{13}CO}$ mixture, respectively. This experiment was performed under a low thickness regime for the water ice, with less than 30 ML of H$_2$O formed. The desorption yields of both CO (represented in red) and $^{13}CO_2$ (in blue) follow the same trend observed in the 3:1 H$_2$O:~CO studies. Notably, the trapped fraction of $^{13}CO_2$ considerably exceeds the value obtained for CO, achieving a value of 0.75 that is out of the range of the plot. This indicates that carbon dioxide has a higher tendency to become trapped within an \textit{in situ}-generated water film compared to a film resulting from molecular beam-deposited water. This could potentially be attributed to the entrapment of $^{13}CO_2$ within water clusters during its synthesis, given that $^{13}$CO$_2$ is also formed \textit{in situ} via chemical reactions from $^{13}$CO.

\section{Discussion}
\label{Discussion}

\subsection{CO-like species: How the trapping of volatiles occurs in compact ices}

The entrapment process of volatile molecules in the amorphous structure of water ice is still not fully understood. The desorption and trapping behaviours of these volatile species are primarily dependent on the morphology of the water ice. To the best of our knowledge, the works of \cite{MayI_JCPA_2013, MayII_JCPA_2013} represent the primary studies to experimentally investigate entrapment in highly compact ices, specifically analysing the effects of crystallisation-induced cracks through the ice. Earlier studies, mainly focused on astrophysical ices \citep[e.g.][]{Collings_APJ03, Collings_ASS03, Collings_2004MNRAS_survey, FayolleAA2011, MartinAA2014}, investigated the entrapment within more porous films.

Regarding entrapment in porous ice, the initial study to thoroughly investigate this process and to propose a qualitative model for the entrapment of volatiles in astrophysical water ice analogues was undertaken by \cite{Collings_APJ03}. Their model proposes that when a CO layer, formed at 10 K on porous water ice, is heated to between 15 and 30 K, desorption of solid CO occurs alongside some diffusion of CO into the H$_2$O film. They suggest that the collapse of the pores between 30 and 70 K, along with ice compaction, is responsible for the entrapment of molecules.

In this model, pore structures and their collapse serve as the predominant mechanisms for trapping molecules. However, certain aspects of the trapping process are not fully explained by pore collapse alone. As pointed out by \cite{FayolleAA2011}, the collapse of pores between 30 - 70 K is not sufficient to explain the entrapment, given that the same behaviour is observed in mixtures of water ice and CO$_2$. Consequently, pore collapse should be equally effective at around 30 K during the desorption of CO and at 70 K when CO$_2$ begins to desorb. 

In the current study, we examine a contrasting case by conducting experiments on a highly compact ice film to explore the dynamics of trapping when porosity is significantly reduced. Within the context of trapping in amorphous, compact water ice, volatile molecules need to be incorporated between clusters of water molecules during co-deposition for successful entrapment. These clusters are isolated from the gas phase, preventing the desorption of the volatile species. 
Moreover, we observe that the compact ice films trap volatile species more effectively than porous ice does. This is because the percolation of molecules into pores allows molecules to reach the gas phase. The desorption yields obtained for compact ice in our study can be compared to the previous values reported by \cite{Malyk2007} and \cite{FayolleAA2011} for porous ice. The desorption yield for the trapping efficiency of water, as shown in Fig. \ref{Fractions}b), indicates that compact water ice can capture up to 20\% of the volatiles deposited, relative to the amount of water ice deposited. In porous ice, these yields for H$_2$O : CO ratios ranging from 1 : 1 to 10 : 1 are approximately 5\%, according to \cite{FayolleAA2011}. The water trapping efficiency for CO is found to be $CO_{\text{eff-water}}\approx$ 0.05, across a variety of experiments with surface coverages ranging from 12 to 32 ML. Experiments by \cite{Malyk2007} demonstrated that ices formed by the co-deposition of H$_2$O and CO$_2$ could retain approximately one CO$_2$ molecule for every 30 H$_2$O molecules, given a water ice coverage of around 40 ML. This corresponds to a water trapping efficiency of $CO_{2~eff-water}\approx$ 0.03.

\begin{figure*}[ht]
   \centering
   \includegraphics[width = 18 cm]{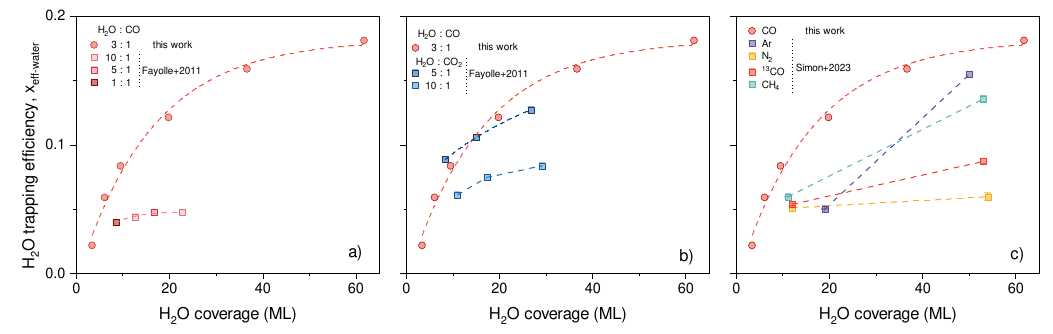}
    \caption{H$_2$O trapping efficiency for volatile species as a function of H$_2$O surface coverage. CO data from this study (circles) and literature (squares). (a) Variation with H$_2$O to CO ratios, highlighting ice matrix porosity effects. (b) Comparable efficiencies for CO and CO$_2$ due to ice structure. (c) Efficiency trends for hypervolatiles, including data on $^{13}$CO, CH$_4$, and N$_2$ from \cite{Simon2023}.}
   \label{Fractions_ref}
    \end{figure*}

To discuss the trapping mechanisms in compact and porous ices,   
in Figure \ref{Fractions_ref} we compare the trapping efficiency of water for different molecules as observed in our experiments of H$_2$O + CO ratio 3 : 1 (indicated by circles) with the data reported in the study by \cite{FayolleAA2011} and \citep{Simon2023} with more porous ices (represented by squares).

Panel (a) shows the trapping efficiency variation with different ratios of H$_2$O to CO. For compact ice, trapping efficiency exhibits a non-linear increase with ice coverage. The rise in efficiency flattens as the coverage approaches 60 ML, suggesting a saturation point.

In contrast, the trapping efficiency for a more porous ice matrix, represented by squares, shows a markedly different behaviour. It remains relatively low and does not display a significant increase with ice coverage, remaining approximately constant, despite the different H$_2$O:CO ratios explored: 10 : 1, 5 : 1, and 1 : 1.  This plot underscores the importance of ice structure in the trapping mechanism, with compact ice showing a significantly higher efficiency compared to porous ice across the measured coverage range.

Panel b) presents the trapping efficiency of H$_2$O for CO, based on the current study, and for CO$_2$, as reported by \cite{FayolleAA2011}. 
The efficiency of water ice trapping CO$_2$ in a porous matrix at the higher H$_2$O:CO$_2$ ratio of 5:1 appears to mirror the pattern observed for H$_2$O trapping CO in a compact ice structure. In contrast, at the lower ratio of 10:1, CO$_2$ demonstrates a decreased trapping efficiency within the porous ice. 

For the H$_2$O : CO$_2$ ratio 5 : 1, the observed behaviour can be attributed to the role of pore collapse in porous films. For CO$_2$, the collapse of pores, occurring between 30 and 70 K, may hinder the desorption pathway, trapping the molecules before the start of the initial CO$_2$ desorption peak at 80 K. As a result, the trapping efficiency of water for CO$_2$ appears higher in porous films and is similar to the one observed in compact ices.

In our experiments using compact amorphous water ice, we observe a greater uniformity in the ice structure. Hence, the disparities in trapping efficiency for CO and CO$_2$ evident in porous films are not observed in our studies. We consistently record the same desorption yields for both CO and CO$_2$ as showed in Fig. \ref{Fractions}e).

The lower trapping efficiency observed for the 10~:~1 H$_2$O:CO$_2$ ratio can be attributed to the reduced presence of CO$_2$ molecules within the ice matrix. Despite this apparent inefficiency, it is important to note that the available CO$_2$ molecules are indeed being trapped effectively within the water matrix. This indicates that while the absolute efficiency is low due to the lesser amount of CO$_2$ relative to H$_2$O, the trapping mechanism itself remains efficient for the CO$_2$ molecules that are present.

Finally, panel c) of Fig. \ref{Fractions_ref} shows recent results of the trapping efficiency of water for other hyper-volatile molecules ($^{13}$CO, CH$_4$, Ar, and N$_2$ from \citep{Simon2023} compared to the trapping efficiency of H$_2$O for CO, based on the current study. 

The data from \cite{Simon2023} show that the trapping efficiencies for hyper-volatile species such as Ar and CH$_4$ increase alongside the H$_2$O coverage. However, this increase occurs at a distinct rate when compared to $^{13}$CO and N$_2$. Despite all four species being hyper-volatile and tending to desorb before the collapse of the pores within the 30 - 70 K temperature range, their trapping efficiencies within the water matrix differ.

As suggested by \cite{MayII_JCPA_2013}, the diffusion of gases through water ice layers is not solely controlled by the pore structure and connected pathways. Instead, it is influenced by factors such as the size of the atoms and molecules and the morphology of the water ice itself. This variance in trapping efficiency could be attributed to the differences in bulk diffusion behaviour of the species. Each type of atom and molecules diffuses through the water matrix at a different rate, which affects how effectively they desorb in the first desorption event and how much is trapped. These findings suggest that while pore collapse in porous ices may be a common factor affecting all hyper-volatile species, the specific interaction between each gas and the water matrix, likely influenced by their individual diffusion characteristics, results in the observed diversity in trapping efficiency trends. These differences indicate the importance studying bulk diffusion processes, as they evidently play a partial role in the trapping behaviour of volatiles. 

\subsection{A view for entrapment in a compact ice}

We propose a simplified view to describe the entrapment in a compact ice including the gas phase, surface, and ice mantle aiming to simulate the observed trends in desorption yields for volatiles seen in Fig. \ref{Fractions}. This attempt resembles to models published by \cite{HasegawaMNRAS93} and later by \cite{FayolleAA2011}, which include kinetics parameters for the desorption such as the binding energies and bulk diffusion. We acknowledge that this view, akin to a 'toy model', is designed to provide a basic physical interpretation and to represent general trends, rather than to be a precise quantitative tool. For a more detailed and comprehensive model, we will present a bench-marking of the LABICE code using the data-set introduced in this initial paper. In the future we aim to underscore the challenges involved in taking into account morphology over time and temperature to model the desorption yields correctly.

\begin{figure}[ht]
   \centering
   \includegraphics[width = 9 cm]{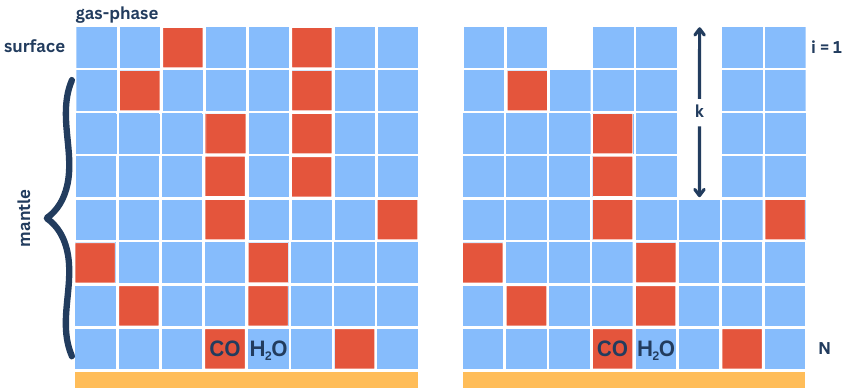}
   \caption{Simplified view to describe the entrapment in a
compact ice adapted from \cite{HasegawaMNRAS93} and \cite{FayolleAA2011}. It shows the ice structure before (left) and after (right) molecules of CO desorb from the surface to the gas phase, with the bulk ice mantle, surface monolayer, and gas phase delineated. \textit{k} is the extend of the pathway left in the ice due to the first peak of CO desorption.}
   \label{sketch}
    \end{figure}

Figure \ref{sketch} depicts a schematic representation of the entrapment process in a compact ice matrix. In total, there are N layers, with layer i = 1 being the layer in contact with the gas phase, and layer i = N being the one that is absorbed by the substrate. Each layer is composed of H$_2$O and CO molecules and the the extent to which CO  molecules are trapped within water ice is determined by the length of the pathways created in the ice during the initial peak of CO desorption. Specifically, the likelihood that a CO molecule will desorb from a given layer is:

\begin{equation}
P_{\text{des}} = \frac{<k>}{N}, \text{ where } <k> = a \cdot \sum_{k=0}^{\infty} k \cdot b^k = \frac{a \cdot b}{(1 - b)^2} 
\label{Probability_des}
\end{equation}

The ratio H$_2$O : CO sets the values of \textit{b}, which represent the amount of CO per layer and \textit{a} is the amount of water per layer. In a binary ice mixture, \textit{a = 1 - b} and the sum of the terms of Eq. \ref{Probability_des} can be written as: 

\begin{equation}
P_{\text{des}} = \frac{b}{N(1 - b)}
\label{Probability_des_2}
\end{equation}

In the layer in contact with the surface, for a ratio H$_2$O : CO = 3 : 1, $P_{des}$ = 0.33/N. As one moves deeper into the ice layers, the probability of desorption decreases, resulting in the entrapment of the volatile molecule. It is important to note that Eq. \ref{Probability_des}, does not directly incorporate temperature, as our 'toy model' is designed to understand the desorption trends rather than to describe the kinetics of the process. It is obviously too simplistic.

\begin{figure}[ht]
   \centering
   \includegraphics[width = 6.0 cm]{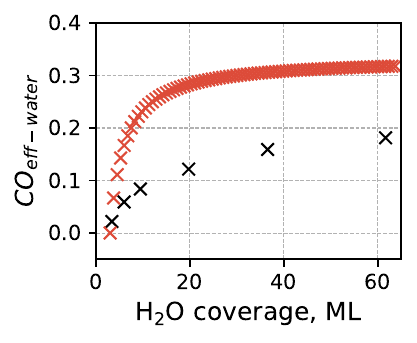}   
   \caption{Reproduced trapping efficiency of water for the CO desorption using equation \ref{Probability_des} compared with experiments of H$_2$O : CO = 3 : 1. The  equation is capable of replicating the plateau as well as the general behaviour of experiments, however, it tends to overestimate the absolute values of the desorption yield.}
   \label{Fig_model}
    \end{figure}

Figure \ref{Fig_model} compares the trapping efficiency of water for CO obtained with this model and the experimental values from the binary mixtures of H$_2$O : CO = 3 : 1 (in black). When the desorption yields are plotted in function of the water surface coverage (Fig. \ref{Fractions}), our experimental data demonstrates that the quantity of volatile molecules trapped in the ice increases with coverage until it reaches a plateau at around 20 ML. Equation \ref{Probability_des} is able to reproduce this curve behaviour, but it overestimates the absolute values of desorption yields. This is expected from a simple model because it does not take into account factors such as the restructuring of the ice due to the increasing temperature and the binding energy resulting from molecular interactions. Nevertheless, it serves as a useful tool in describing the mechanical entrapment of hyper volatile molecules within compact ices.

We point out that there is no fitting or free parameter, and our aim is not to accurately fit our data, that we could do by adjusting the height of the plateau. Our aim is to put light on the role of exogenous molecules in the formation and frame of the water ice. By definition a mixed film incorporate both water and an exogenous molecules, and while water is a priory made compact in our experimental conditions, the presence of exogenous molecules actually creates lacunae that limits the intrinsic capability of water to trap theses molecules. Therefore we demonstrate that in this very thin layers conditions, water ice cannot retain volatile molecules, at least not more than the limit of the plateau of the trapping efficiency, which is about 0.2. At least, five water molecules are required to store one exogenous molecule.

\subsection{H$_2$O-like specie}

The entrapment mechanisms for water-like species and highly volatile molecules exhibit differences. For H-bonded species, entrapment can involve processes like the formation of clathrate structures, in which water molecules organise into 'cages' capable of confining other molecules, such as methanol \citep{Notesco2000}.

In terms of the desorption yields, section \ref{des_yield_metOH} demonstrates that the trapping efficiency of water for methanol remain consistent across varying ice structures. Methanol molecules stay entrapped in the water ice up to a concentration of about 3\%, desorbing in the co-desorption peak. Beyond a concentration of approximately 3\%, methanol will exhibit a desorption peak equivalent to that of pure methanol prior to the co-desorption peak. Similar findings were observed in experiments conducted by \cite{MartinAA2014}, which reported a water trapping efficiency of 3\% for methanol and 5\% for ammonia. These values remained constant for both molecules in cometary-analogue ices, regardless of thickness variations ranging from 300 to 800 ML.

\subsection{Ternary mixtures}

Multi-component ice mixtures often display complex behaviours during desorption, particularly when the mixtures consist of molecules from diverse desorption categories, such as volatiles and H$_2$O-like species. In experiments focusing on ternary mixtures of H$_2$O, CH$_3$OH, and CO, findings indicate that methanol does not impact the trapping efficiency of water for CO, at least within the tested surface coverage range of 20 ML of water. However, the presence of methanol does influence the structural properties of the ice. It specifically affects the crystallisation of water, resulting in a less well-defined volcano peak compared to binary mixtures that do not contain methanol.

The alteration in crystallisation kinetics is a consequence of the interaction between methanol and amorphous water ice. Methanol, when mixed with the ice, can promote the formation of ice crystals through a mechanism known as heterogeneous nucleation \citep{SoudaPhysRevB, Tonauer2023}. This process lower the energy barrier required for crystal formation, thereby enhancing the nucleation rate and consequently altering the crystallisation kinetics. Such a modification is evident in the lowered temperature required for crystallisation, as corroborated by TPD curves and the absence of the characteristic crystallisation 'bump'.

\subsection{H$_2$O ice formed \textit{in situ}}

The structure of water ices in astrophysical environments continues to be a topic of debate. When solid water is deposited in a laboratory setting at low temperatures (around 10 K) and slow adsorption rates, the film is formed through a process known as ballistic deposition, where molecules "hit and stick" to the surface \citep{Kimmel2001JChPh_sim, KimmelJChPh2001_exp}. This produces a water ice with a porous, amorphous structure of low density. Given that dark clouds exhibit similar conditions (low temperatures, slow adsorption rates, and random paths of adsorbate molecules), it has been suggested by numerous authors that water ices accumulating in these environments would be porous and amorphous. In this context, water would only become compact when pores merge during thermal processing at temperatures above 30~K. Nonetheless, as most water in dark clouds forms on the surface of dust grains rather than accreating from the gas phase, the predicted structure of water ices originating in dark clouds is that of a dense and amorphous solid \citep{Minissale_review_22}. Laboratory-generated water ice analogues formed \textit{in situ} from hydrogen and oxygen atomic beams exhibit a compact structure as opposed to a porous one, regardless of the deposition angle. The formation of dense ice is facilitated by the enthalpy release of the water formation reactions, causing localised heating that drives the compaction \citep{Accolla_MNRAS_2013}.
In the present study, we demonstrate that compact water ice formed \textit{in situ}, which as we discussed can also be viewed as a more accurate model of interstellar ice, particularly in dark cloud environments, is capable of trapping volatiles with an efficiency equal to that of the compact water deposited. This highlights the importance of determining the morphology of interstellar ices, as it will provide insights into how to model essential processes in the solid phase.

\subsection{The effect of different heating rates}

The heating rate is a critical parameter in transposing laboratory experiments to astrophysical environments, and it is important to discuss the implications of our chosen heating rates in this context, specifically within the ISM, and the effect different heating rates could have on the trapping efficiencies presented in this study. 

In the ISM, volatile loss from ices occurs as dust grains are heated. A difference between experimental conditions and the ISM could arise if grains in the ISM are primarily heated from the outside, leading to a surface-to-core warming profile. This would contrast with laboratory settings where substrate-to-surface heating predominates, potentially reversing the temperature gradient across the ice. Despite these differences, the heating rates used in our experiments are designed to simulate slow heating processes that are astrophysically relevant. Given the small size of ISM  grains, their rotational dynamics, and the heating influence of the protostar, the heating process at the grain scale is likely more homogeneous and isotropic. On the other hand, we do not expect to have a relevant temperature gradient in the laboratory experiments due to the slow heating rates adopted and the low thickness of the ices studied.

In this work, we also adopted a uniform heating rate of 0.2 K/s in the dataset. Choosing different heating rates can result in shifts in peak temperatures and changes in the shape of the desorption peaks. In the ISM, assuming a heating ramp of 1 K/century would simply shift the desorption to much lower temperatures. Conversely, rapid heating ramps could potentially cause bursts of gas release, driven by gas pressure from below forcing the surface water substrate outward. This phenomenon is observed in the experiments conducted by \cite{MayII_JCPA_2013}, where a multilayer of different high-volatile gases is deposited underneath water ice layers. For more homogeneous mixtures, which are more realistic scenarios for ISM ices, the gas pressure would be less pronounced to cause the rupture of the ice and a burst of volatile desorption.

In addition, the effect of slow ramp rates must be discussed to understand if such rates would permit enough time for the diffusion of gases within the water ice, resulting in all the gas being released at low temperatures. Supporting this, the studies by \cite{MayI_JCPA_2013, MayII_JCPA_2013} conducted an in-depth analysis of the effect of heating rate on gas trapping across different surface coverage regimes. Notably, the study reveals that the low-temperature peaks in the desorption spectra are not solely due to diffusion, indicating that gases will still be trapped even with a very slow heating ramp. This is also supported by previous studies from \cite{Mispelaer2013AA} and \cite{GhesquiereAA18} suggesting that the release of gases due to diffusion has a minimal impact once the ice layer is compact.

\section{Astrophysical relevance}
\label{conclusions}

In order to model the organic content and solid-state chemistry inherent in interstellar ices, several processes must be considered: gas-phase accretion, surface and ice mantle diffusion of atoms, radicals, and molecules, the reactivity of species upon encounter, and both thermal and non-thermal desorption. The desorption yields reported in this paper are especially pertinent in understanding the interplay between bulk diffusion and reactivity in the ice mantle, which can lead to the formation of interstellar complex organic molecules (iCOMs).

The dynamics of the chemistry involved in iCOMs formation, in the context of low-temperature and out-of-equilibrium environments, are dictated by the reaction-diffusion equation. This equation delineates the Langmuir-Hinshelwood mechanism on surfaces and within the ice mantle, and is reliant on temperature-dependent reaction rate constants and diffusion coefficients. The time it takes for the reaction and diffusion to take place influences the kinetics and yield of the reaction. The reaction-diffusion process within a multi-layer ice is primarily governed by the restructuring of the water ice mantle \citep{Mispelaer2013AA, GhesquiereAA18}, while the availability of the species for high-temperature reactions is dictated by their desorption yields.

The reactivity mechanism driven by structural changes in the bulk ice bears significant astrophysical implications, as it enables radical-radical, radical-molecule, and molecule-molecule reactions to occur at a rate significantly higher than the previously expected bulk diffusion rate. This is especially relevant within star-forming regions where ices may be heated to temperatures facilitating both rotational (T < 115 K) and long-range translational (T < 120 K) diffusion of heavy species. Under these conditions, the molecules or radicals observed in the IR spectra of ices can effectively form iCOMs within the volume of the icy mantles. This aspect must be taken into account when modelling iCOMs observations \citep{Mispelaer2013AA, GhesquiereAA18, Theule2020IAUS}.

The findings from this study regarding the trapping efficiency of water offer insights into the proportion of species expected to be available for reaction, thus shedding light on the limited number of iCOMs that can be generated upon mantle warm-up. Our experiments illustrates that compact water ices are capable of trapping a variety of molecules: volatiles, H-bonding molecules, and semi-refractory species. However, the overall trapping capacity is finite. In compact water ices, volatile molecules are captured between the amorphous ice clusters, with the free path for desorption only accessible for molecules within the initial ice layers proximate to the gas phase. Pores mitigates desorption yield, showing that compact water ices can trap more species than porous films.

Our results confirm that even compact ices have a maximal trapping capacity of 20\% for volatiles (Ar, CO, CO$_2$), 3 - 5 \% for H-bonding molecules (CH$_3$OH, NH$_3$), and zero for semi-refractory species (NH$_4$$^+$HCOO$^-$). It is important to note that our studies primarily involved closed-shell molecules due to the experimental challenges associated with radicals. Nevertheless, these trapping efficiencies can be extrapolated to estimate the quantity of radicals that could potentially be trapped within the ice mantle and be available for reactions. For instance, the trapping efficiency of water for methanol, which stands at 3\%, could be extrapolated to the availability of CH$^3$O$^{.}$ radicals for reactions within the mantle. 

Besides, the semi-volatile character of species such as ammonium salts, exemplifies the importance of accurately modelling the thermal desorption behaviour in accordance with their respective desorption categories \citep{KruczkiewiczAA}. The semi-volatile nature of some ammonium salts could explain the missing nitrogen in the gas-phase content in comets, as showed by data from the Rosetta mission \citep{AltweggNatA20, PochScience20, Altwegg2022}.

Recent JWST measurements of ices in dense molecular cloud indicate that CO and CO$_2$ exhibit an abundance approximately 20\% relative to water \citep{McClure_JWST}. However, the retention of these molecules within the ice and their potential participation in subsequent reactions during advanced stages of star formation depends on a variety of factors, as outlined in this paper. These factors comprise the degree of mixing or segregation of molecules in a multilayered structure, and the compactness or porosity of the ice. The trapping efficiency can vary between highly compact and porous ices, resulting in disparate trapping behaviours for volatile species. As such, understanding the morphology of interstellar water ice is crucial for determining the availability of these species for modelling subsequent process during star formation.

\section{Conclusions}

In the present work, we carried out ice sublimation experiments with the goal of offering data for astrochemical models. We conducted experiments with increasing order of complexity (pure, binary, and ternary mixtures) on compact amorphous water ice films and with different species (Ar, CO, CO$_2$, NH$_3$, CH$_3$OH, NH$_4$$^+$HCOO$^-$). From the desorption yields, we find common trends in the trapping of molecules when their abundance is compared to water. Our main findings can be summarised as follows: 

\begin{enumerate}
    \item Within binary ice mixtures, water ice is capable of trapping up to 20\% of volatile molecules (Ar, CO, and CO$_2$), with this percentage reflecting the proportion of these molecules relative to the total abundance of water ice deposited. Hydrogen-bonded molecules like CH$_3$OH and NH$_3$ are trapped with lower efficiencies, approximately 3\% for methanol and 5\% for ammonia, whereas ammonium formate (NH$_4$$^+$HCOO$^-$) shows no significant volcano or co-desorption with the water ice matrix.
    \item Our analysis of ternary ice mixtures indicates that the entrapment efficiencies for highly volatile species remain consistent, as exemplified in the H$_2$O:CH$_3$OH:CO system. The presence of methanol influences the crystallisation kinetics of water, although it does not notably alter the entrapment efficiency of the volatile components.
    \item We find that compact amorphous water ice, whether formed in situ or deposited through molecular beams, exhibits comparable trapping capabilities. Specifically, in co-deposition experiments involving a mixture of {O$_2$} + H + $^{13}$CO, we observe the concurrent production of water via hydrogenation of molecular oxygen, with $^{13}$CO being trapped efficiently. We also note the formation of $^{13}$CO$_2$ and $^{13}$CH$_3$OH through hydrogenation processes as well as their entrapment within the water ice. 
    \item Our results indicate that compact water ices show a greater trapping capacity than porous ices. Nonetheless, the trapping efficiency of CO$_2$ within porous ices—which desorbs during the restructuring phase of the ice between 30 and  70 K—is equal to that within compact ices, demonstrating the role of pore collapse in the entrapment process within porous matrices.
\end{enumerate}

These desorption yields have implications for predicting the volatile abundances within protoplanetary discs and, by extension, the composition of comets. Therefore, incorporating detailed entrapment processes into astrochemical models is essential for an accurate representation of chemical abundances and interactions in planet-forming regions. Additionally, the desorption yields provide insights into the availability of species to react and form iCOMs during the warm-up phase of ice mantles.

Additionally, we provide a comprehensive and step-by-step set of experimental data that can be used to benchmark gas-grain astrochemical models. To this end, we  propose the following step-by-step method:

\begin{enumerate}
    \item Begin by using the desorption parameters listed in Table \ref{<Edes>} to reproduce the pure-ice TPD curves shown in Fig.\ref{Fig_H2O_CO}(a) (with the crystallisation bump) and (b), Fig. \ref{Fig_H2O_Ar}(a), Fig. \ref{Fig_H2O_CO2}(a), Fig. \ref{Fig_H2O_NH3}(a), Fig. \ref{Fig_H2O_CH3OH}(a), and Fig. \ref{Fig_H2O_NH4HCOO}(a). These figures provide examples of the desorption of a representative from each category of molecules, as detailed in Section \ref{species_selection}. This initial step helps to ensure the appropriate use of desorption parameters and the correct representation of water crystallisation.
    \item Proceed to reproduce the binary ice mixtures shown in Fig.\ref{Fig_H2O_CO}(c), Fig. \ref{Fig_H2O_Ar}(b), Fig. \ref{Fig_H2O_CO2}(b), Fig. \ref{Fig_H2O_NH3}(b), Fig. \ref{Fig_H2O_CH3OH}(b), and Fig. \ref{Fig_H2O_NH4HCOO}(b). Both the kinetics (the positioning of the desorption peaks) and desorption yields (the ratios of the integrated desorption peaks) should align with the data in Table \ref{table_overview}.
    \item  Lastly, attempt to reproduce more complex mixtures, such as ternary and multi-component ice mixtures as shown in Fig.\ref{Fig_H2O_CO}(c) and Fig.\ref{Fig_H2Oinsitu_CH3OH_X}, taking into account both kinetics and desorption yields. These mixtures are more representative of actual interstellar ices. This final step should be feasible if the two previous steps have been successfully completed.
\end{enumerate}

In a forthcoming paper, we will present a benchmarking of the LABICE code using the dataset introduced in this initial paper. We aim to underscore the challenges involved in taking into account morphology over time and temperature when modelling the desorption yields correctly. Benchmarking models of desorption and molecular segregation based on laboratory experiments is a critical prerequisite for achieving a reliable description of phenomena such as snow lines, shock regions, and comet outgassing.

\begin{acknowledgements}

This project has greatly benefited from various funding sources. Primarily, we would like to acknowledge the support received from the European Union’s Horizon 2020 research and innovation program, specifically under the Marie Skłodowska-Curie grant agreement No $811312$ assigned for the "Astro-Chemical Origins" (ACO) project. Additionally, we appreciate the backing provided by the ANR SIRC project (GrantANRSPV202448 2020-2024). We also gratefully recognise the support from the National Programme "Physique et Chimie du Milieu Interstellaire" (PCMI) of CNRS/INSU, co-funded by CEA and CNES, and the DIM ACAV+, a funding initiative by the Region Ile de France. Lastly, we would like to extend our sincerest gratitude to the Max Planck Society for their continual support.

\end{acknowledgements}

\bibliographystyle{aa} 
\bibliography{REFs}

\begin{thebibliography}{104}
\expandafter\ifx\csname natexlab\endcsname\relax\def\natexlab#1{#1}\fi

\bibitem[{{Accolla} {et~al.}(2011){Accolla}, {Congiu}, {Dulieu}, {Manic{\`o}}, {Chaabouni}, {Matar}, {Mokrane}, {Lemaire}, \& {Pirronello}}]{Accolla2011}
{Accolla}, M., {Congiu}, E., {Dulieu}, F., {et~al.} 2011, Physical Chemistry Chemical Physics (Incorporating Faraday Transactions), 13, 8037

\bibitem[{{Accolla} {et~al.}(2013){Accolla}, {Congiu}, {Manic{\`o}}, {Dulieu}, {Chaabouni}, {Lemaire}, \& {Pirronello}}]{Accolla_MNRAS_2013}
{Accolla}, M., {Congiu}, E., {Manic{\`o}}, G., {et~al.} 2013, \mnras, 429, 3200

\bibitem[{{Altwegg} {et~al.}(2020){Altwegg}, {Balsiger}, {H{\"a}nni}, {Rubin}, {Schuhmann}, {Schroeder}, {S{\'e}mon}, {Wampfler}, {Berthelier}, {Briois}, {Combi}, {Gombosi}, {Cottin}, {De Keyser}, {Dhooghe}, {Fiethe}, \& {Fuselier}}]{AltweggNatA20}
{Altwegg}, K., {Balsiger}, H., {H{\"a}nni}, N., {et~al.} 2020, Nat. Astron., 3

\bibitem[{{Altwegg} {et~al.}(2022){Altwegg}, {Combi}, {Fuselier}, {H{\"a}nni}, {De Keyser}, {Mahjoub}, {M{\"u}ller}, {Pestoni}, {Rubin}, \& {Wampfler}}]{Altwegg2022}
{Altwegg}, K., {Combi}, M., {Fuselier}, S.~A., {et~al.} 2022, \mnras, 516, 3900

\bibitem[{{Biham} {et~al.}(2001){Biham}, {Furman}, {Pirronello}, \& {Vidali}}]{BihamApJ01}
{Biham}, O., {Furman}, I., {Pirronello}, V., \& {Vidali}, G. 2001, The Astrophysical Journal, 553, 595

\bibitem[{{Bolina} \& {Brown}(2005)}]{BolinaSurSc05}
{Bolina}, A.~S. \& {Brown}, W.~A. 2005, Surface Science, 598, 45

\bibitem[{{Bolina} {et~al.}(2005){Bolina}, {Wolff}, \& {Brown}}]{BolinaJCP05}
{Bolina}, A.~S., {Wolff}, A.~J., \& {Brown}, W.~A. 2005, Journal of Chemical Physics, 122, 044713

\bibitem[{{Boogert} {et~al.}(2015){Boogert}, {Gerakines}, \& {Whittet}}]{BoogertARAA15}
{Boogert}, A. C.~A., {Gerakines}, P.~A., \& {Whittet}, D. C.~B. 2015, Annual Review of Astronomy and Astrophysics, 53, 541

\bibitem[{{Burke} \& {Brown}(2015)}]{BurkeBrown15}
{Burke}, D.~J. \& {Brown}, W.~A. 2015, Monthly Notices of the Royal Astronomical Society, 448, 1807

\bibitem[{{Carter}(1962)}]{CarterVacuum62}
{Carter}, G. 1962, Vacuum, 12, 245

\bibitem[{{Cazaux} {et~al.}(2003){Cazaux}, {Tielens}, {Ceccarelli}, {Castets}, {Wakelam}, {Caux}, {Parise}, \& {Teyssier}}]{Cazaux2003}
{Cazaux}, S., {Tielens}, A.~G.~G.~M., {Ceccarelli}, C., {et~al.} 2003, \apjl, 593, L51

\bibitem[{{Ceccarelli}(2004)}]{Ceccarelli2004}
{Ceccarelli}, C. 2004, in Astronomical Society of the Pacific Conference Series, Vol. 323, Star Formation in the Interstellar Medium: In Honor of David Hollenbach, ed. D.~{Johnstone}, F.~C. {Adams}, D.~N.~C. {Lin}, D.~A. {Neufeeld}, \& E.~C. {Ostriker}, 195

\bibitem[{{Ceccarelli} {et~al.}(2017){Ceccarelli}, {Caselli}, {Fontani}, {Neri}, {L{\'o}pez-Sepulcre}, {Codella}, {Feng}, {Jim{\'e}nez-Serra}, {Lefloch}, {Pineda}, {Vastel}, {Alves}, {Bachiller}, {Balucani}, {Bianchi}, {Bizzocchi}, {Bottinelli}, {Caux}, {Chac{\'o}n-Tanarro}, {Choudhury}, {Coutens}, {Dulieu}, {Favre}, {Hily-Blant}, {Holdship}, {Kahane}, {Jaber Al-Edhari}, {Laas}, {Ospina}, {Oya}, {Podio}, {Pon}, {Punanova}, {Quenard}, {Rimola}, {Sakai}, {Sims}, {Spezzano}, {Taquet}, {Testi}, {Theul{\'e}}, {Ugliengo}, {Vasyunin}, {Viti}, {Wiesenfeld}, \& {Yamamoto}}]{Ceccarelli2017}
{Ceccarelli}, C., {Caselli}, P., {Fontani}, F., {et~al.} 2017, \apj, 850, 176

\bibitem[{{Ceccarelli} {et~al.}(2023){Ceccarelli}, {Codella}, {Balucani}, {Bockelee-Morvan}, {Herbst}, {Vastel}, {Caselli}, {Favre}, {Lefloch}, {Oberg}, \& {Yamamoto}}]{Ceccarelli2022}
{Ceccarelli}, C., {Codella}, C., {Balucani}, N., {et~al.} 2023, in Astronomical Society of the Pacific Conference Series, Vol. 534, Protostars and Planets VII, ed. S.~{Inutsuka}, Y.~{Aikawa}, T.~{Muto}, K.~{Tomida}, \& M.~{Tamura}, 379

\bibitem[{{Chahine} {et~al.}(2022){Chahine}, {L{\'o}pez-Sepulcre}, {Neri}, {Ceccarelli}, {Mercimek}, {Codella}, {Bouvier}, {Bianchi}, {Favre}, {Podio}, {Alves}, {Sakai}, \& {Yamamoto}}]{Chahine2022}
{Chahine}, L., {L{\'o}pez-Sepulcre}, A., {Neri}, R., {et~al.} 2022, \aap, 657, A78

\bibitem[{{Chang} {et~al.}(2007){Chang}, {Cuppen}, \& {Herbst}}]{ChangAA07}
{Chang}, Q., {Cuppen}, H.~M., \& {Herbst}, E. 2007, \aap, 469, 973

\bibitem[{{Chuang} {et~al.}(2016){Chuang}, {Fedoseev}, {Ioppolo}, {van Dishoeck}, \& {Linnartz}}]{Chuang2016}
{Chuang}, K.~J., {Fedoseev}, G., {Ioppolo}, S., {van Dishoeck}, E.~F., \& {Linnartz}, H. 2016, \mnras, 455, 1702

\bibitem[{{Collings} {et~al.}(2004){Collings}, {Anderson}, {Chen}, {Dever}, {Viti}, {Williams}, \& {McCoustra}}]{Collings_2004MNRAS_survey}
{Collings}, M.~P., {Anderson}, M.~A., {Chen}, R., {et~al.} 2004, \mnras, 354, 1133

\bibitem[{{Collings} {et~al.}(2003{\natexlab{a}}){Collings}, {Dever}, {Fraser}, \& {McCoustra}}]{Collings_ASS03}
{Collings}, M.~P., {Dever}, J.~W., {Fraser}, H.~J., \& {McCoustra}, M.~R.~S. 2003{\natexlab{a}}, \apss, 285, 633

\bibitem[{{Collings} {et~al.}(2003{\natexlab{b}}){Collings}, {Dever}, {Fraser}, {McCoustra}, \& {Williams}}]{Collings_APJ03}
{Collings}, M.~P., {Dever}, J.~W., {Fraser}, H.~J., {McCoustra}, M.~R.~S., \& {Williams}, D.~A. 2003{\natexlab{b}}, \apj, 583, 1058

\bibitem[{Congiu {et~al.}(2020)Congiu, Sow, Nguyen, Baouche, \& Dulieu}]{Congiu_RSI_2020}
Congiu, E., Sow, A., Nguyen, T., Baouche, S., \& Dulieu, F. 2020, Review of Scientific Instruments, 91, 124504

\bibitem[{{Cooke} {et~al.}(2018){Cooke}, {{\"O}berg}, {Fayolle}, {Peeler}, \& {Bergner}}]{Cooke_APJ_18}
{Cooke}, I.~R., {{\"O}berg}, K.~I., {Fayolle}, E.~C., {Peeler}, Z., \& {Bergner}, J.~B. 2018, \apj, 852, 75

\bibitem[{{Dartois}(2005)}]{Dartois2005}
{Dartois}, E. 2005, \ssr, 119, 293

\bibitem[{{d'Hendecourt} {et~al.}(1985){d'Hendecourt}, {Allamandola}, \& {Greenberg}}]{dHendecourtAA85}
{d'Hendecourt}, L.~B., {Allamandola}, L.~J., \& {Greenberg}, J.~M. 1985, \aap, 152, 130

\bibitem[{{Doronin} {et~al.}(2015){Doronin}, {Bertin}, {Michaut}, {Philippe}, \& {Fillion}}]{DoroninJCP15}
{Doronin}, M., {Bertin}, M., {Michaut}, X., {Philippe}, L., \& {Fillion}, J.~H. 2015, \jcp, 143, 084703

\bibitem[{{Dulieu} {et~al.}(2005){Dulieu}, {Amiaud}, {Baouche}, {Momeni}, {Fillion}, \& {Lemaire}}]{Dulieu2005}
{Dulieu}, F., {Amiaud}, L., {Baouche}, S., {et~al.} 2005, Chemical Physics Letters, 404, 187

\bibitem[{{Fayolle} {et~al.}(2011){Fayolle}, {{\"O}berg}, {Cuppen}, {Visser}, \& {Linnartz}}]{FayolleAA2011}
{Fayolle}, E.~C., {{\"O}berg}, K.~I., {Cuppen}, H.~M., {Visser}, R., \& {Linnartz}, H. 2011, \aap, 529, A74

\bibitem[{{Fedoseev} {et~al.}(2015){Fedoseev}, {Cuppen}, {Ioppolo}, {Lamberts}, \& {Linnartz}}]{Fedoseev2015}
{Fedoseev}, G., {Cuppen}, H.~M., {Ioppolo}, S., {Lamberts}, T., \& {Linnartz}, H. 2015, \mnras, 448, 1288

\bibitem[{{Ferland} {et~al.}(2017){Ferland}, {Chatzikos}, {Guzm{\'a}n}, {Lykins}, {van Hoof}, {Williams}, {Abel}, {Badnell}, {Keenan}, {Porter}, \& {Stancil}}]{FerlandRMxAA17}
{Ferland}, G.~J., {Chatzikos}, M., {Guzm{\'a}n}, F., {et~al.} 2017, \rmxaa, 53, 385

\bibitem[{{Ferrero} {et~al.}(2023){Ferrero}, {Pantaleone}, {Ceccarelli}, {Ugliengo}, {Sodupe}, \& {Rimola}}]{Ferrero2022}
{Ferrero}, S., {Pantaleone}, S., {Ceccarelli}, C., {et~al.} 2023, \apj, 944, 142

\bibitem[{{Ferrero} {et~al.}(2020){Ferrero}, {Zamirri}, {Ceccarelli}, {Witzel}, {Rimola}, \& {Ugliengo}}]{FerreroApJ2020}
{Ferrero}, S., {Zamirri}, L., {Ceccarelli}, C., {et~al.} 2020, \apj, 904, 11

\bibitem[{{Garrod}(2008)}]{GarrodAAP08}
{Garrod}, R.~T. 2008, \aap, 491, 239

\bibitem[{{Garrod}(2013)}]{GarrodApJ13}
{Garrod}, R.~T. 2013, \apj, 778, 158

\bibitem[{{Garrod} \& {Herbst}(2006)}]{Garrod_Herbst2006}
{Garrod}, R.~T. \& {Herbst}, E. 2006, \aap, 457, 927

\bibitem[{{Garrod} {et~al.}(2022){Garrod}, {Jin}, {Matis}, {Jones}, {Willis}, \& {Herbst}}]{Garrod2022ApJS}
{Garrod}, R.~T., {Jin}, M., {Matis}, K.~A., {et~al.} 2022, \apjs, 259, 1

\bibitem[{{Garrod} \& {Pauly}(2011)}]{GarrodApJ11}
{Garrod}, R.~T. \& {Pauly}, T. 2011, \apj, 735, 15

\bibitem[{{Ghesqui{\`e}re} {et~al.}(2018){Ghesqui{\`e}re}, {Ivlev}, {Noble}, \& {Theul{\'e}}}]{GhesquiereAA18}
{Ghesqui{\`e}re}, P., {Ivlev}, A., {Noble}, J.~A., \& {Theul{\'e}}, P. 2018, \aap, 614, A107

\bibitem[{{Green} {et~al.}(2001){Green}, {Toniazzo}, {Pilling}, {Ruffle}, {Bell}, \& {Hartquist}}]{GreenAA01}
{Green}, N.~J.~B., {Toniazzo}, T., {Pilling}, M.~J., {et~al.} 2001, \aap, 375, 1111

\bibitem[{{Harsono} {et~al.}(2015){Harsono}, {Bruderer}, \& {van Dishoeck}}]{HarsonoAA15}
{Harsono}, D., {Bruderer}, S., \& {van Dishoeck}, E.~F. 2015, \aap, 582, A41

\bibitem[{{Hasegawa} \& {Herbst}(1993)}]{HasegawaMNRAS93}
{Hasegawa}, T.~I. \& {Herbst}, E. 1993, \mnras, 263, 589

\bibitem[{{Hasegawa} {et~al.}(1992){Hasegawa}, {Herbst}, \& {Leung}}]{HasegawaApJS92}
{Hasegawa}, T.~I., {Herbst}, E., \& {Leung}, C.~M. 1992, \apjs, 82, 167

\bibitem[{{Herbst}(2014)}]{HerbstPCCP14}
{Herbst}, E. 2014, Physical Chemistry Chemical Physics (Incorporating Faraday Transactions), 16, 3344

\bibitem[{{Herbst} \& {van Dishoeck}(2009)}]{Herbst2009ARA}
{Herbst}, E. \& {van Dishoeck}, E.~F. 2009, \araa, 47, 427

\bibitem[{Hiden(2023)}]{HidenAnalytical2023}
Hiden. 2023, Relative Sensitivity: RS Measurements of Gases, Application Note 282, Hiden Analytical Ltd, available at \url{https://www.hiden.de/wp-content/uploads/pdf/RS_Measurement_of_Gases_-_Hiden_Analytical_App_Note_282.pdf}

\bibitem[{{Holdship} {et~al.}(2017){Holdship}, {Viti}, {Jim{\'e}nez-Serra}, {Makrymallis}, \& {Priestley}}]{HoldshipAJ17}
{Holdship}, J., {Viti}, S., {Jim{\'e}nez-Serra}, I., {Makrymallis}, A., \& {Priestley}, F. 2017, \aj, 154, 38

\bibitem[{{Ioppolo} {et~al.}(2011){Ioppolo}, {van Boheemen}, {Cuppen}, {van Dishoeck}, \& {Linnartz}}]{Ioppolo2011}
{Ioppolo}, S., {van Boheemen}, Y., {Cuppen}, H.~M., {van Dishoeck}, E.~F., \& {Linnartz}, H. 2011, \mnras, 413, 2281

\bibitem[{{Itikawa}(2017)}]{Itikawa2017}
{Itikawa}, Y. 2017, Journal of Physical and Chemical Reference Data, 46, 043103

\bibitem[{{Jenniskens} \& {Blake}(1994)}]{JenniskensScience94}
{Jenniskens}, P. \& {Blake}, D.~F. 1994, Science, 265, 753

\bibitem[{Kakkenpara~Suresh {et~al.}(2024)Kakkenpara~Suresh, Dulieu, Vitorino, \& Caselli}]{Suresh2023}
Kakkenpara~Suresh, S., Dulieu, F., Vitorino, J., \& Caselli, P. 2024, \aap, 682, A163

\bibitem[{{Kimmel} {et~al.}(2001{\natexlab{a}}){Kimmel}, {Dohn{\'a}lek}, {Stevenson}, {Smith}, \& {Kay}}]{Kimmel2001JChPh_sim}
{Kimmel}, G.~A., {Dohn{\'a}lek}, Z., {Stevenson}, K.~P., {Smith}, R.~S., \& {Kay}, B.~D. 2001{\natexlab{a}}, \jcp, 114, 5295

\bibitem[{{Kimmel} {et~al.}(2001{\natexlab{b}}){Kimmel}, {Stevenson}, {Dohn{\'a}lek}, {Smith}, \& {Kay}}]{KimmelJChPh2001_exp}
{Kimmel}, G.~A., {Stevenson}, K.~P., {Dohn{\'a}lek}, Z., {Smith}, R.~S., \& {Kay}, B.~D. 2001{\natexlab{b}}, \jcp, 114, 5284

\bibitem[{Kouchi \& Yamamoto(1995)}]{Kouchi1995}
Kouchi, A. \& Yamamoto, T. 1995, Progress in Crystal Growth and Characterization of Materials, 30, 83

\bibitem[{{Kruczkiewicz} {et~al.}(2021){Kruczkiewicz}, {Vitorino}, {Congiu}, {Theul{\'e}}, \& {Dulieu}}]{KruczkiewiczAA}
{Kruczkiewicz}, F., {Vitorino}, J., {Congiu}, E., {Theul{\'e}}, P., \& {Dulieu}, F. 2021, \aap, 652, A29

\bibitem[{{Lauck} {et~al.}(2015){Lauck}, {Karssemeijer}, {Shulenberger}, {Rajappan}, {{\"O}berg}, \& {Cuppen}}]{Lauck2015}
{Lauck}, T., {Karssemeijer}, L., {Shulenberger}, K., {et~al.} 2015, \apj, 801, 118

\bibitem[{{Le Petit} {et~al.}(2006){Le Petit}, {Nehm{\'e}}, {Le Bourlot}, \& {Roueff}}]{LePetitApJS06}
{Le Petit}, F., {Nehm{\'e}}, C., {Le Bourlot}, J., \& {Roueff}, E. 2006, \apjs, 164, 506

\bibitem[{{Ligterink} \& {Minissale}(2023)}]{Ligterink2023}
{Ligterink}, N.~F.~W. \& {Minissale}, M. 2023, \aap, 676, A80

\bibitem[{Luna {et~al.}(2015)Luna, Millán, Domingo, Santonja, \& Satorre}]{LunaVacuum15}
Luna, R., Millán, C., Domingo, M., Santonja, C., \& Satorre, M. 2015, Vacuum, 122, 154

\bibitem[{{Malyk} {et~al.}(2007){Malyk}, {Kumi}, {Reisler}, \& {Wittig}}]{Malyk2007}
{Malyk}, S., {Kumi}, G., {Reisler}, H., \& {Wittig}, C. 2007, Journal of Physical Chemistry A, 111, 13365

\bibitem[{{Mart{\'\i}n-Dom{\'e}nech} {et~al.}(2015){Mart{\'\i}n-Dom{\'e}nech}, {Manzano-Santamar{\'\i}a}, {Mu{\~n}oz Caro}, {Cruz-D{\'\i}az}, {Chen}, {Herrero}, \& {Tanarro}}]{Martin2015}
{Mart{\'\i}n-Dom{\'e}nech}, R., {Manzano-Santamar{\'\i}a}, J., {Mu{\~n}oz Caro}, G.~M., {et~al.} 2015, \aap, 584, A14

\bibitem[{{Mart{\'\i}n-Dom{\'e}nech} {et~al.}(2014){Mart{\'\i}n-Dom{\'e}nech}, {Mu{\~n}oz Caro}, {Bueno}, \& {Goesmann}}]{MartinAA2014}
{Mart{\'\i}n-Dom{\'e}nech}, R., {Mu{\~n}oz Caro}, G.~M., {Bueno}, J., \& {Goesmann}, F. 2014, \aap, 564, A8

\bibitem[{May {et~al.}(2013{\natexlab{a}})May, Smith, \& Kay}]{MayI_JCPA_2013}
May, R.~A., Smith, R.~S., \& Kay, B.~D. 2013{\natexlab{a}}, The Journal of Chemical Physics, 138

\bibitem[{May {et~al.}(2013{\natexlab{b}})May, Smith, \& Kay}]{MayII_JCPA_2013}
May, R.~A., Smith, R.~S., \& Kay, B.~D. 2013{\natexlab{b}}, The Journal of Chemical Physics, 138

\bibitem[{{McClure} {et~al.}(2023){McClure}, {Rocha}, {Pontoppidan}, {Crouzet}, {Chu}, {Dartois}, {Lamberts}, {Noble}, {Pendleton}, {Perotti}, {Qasim}, {Rachid}, {Smith}, {Sun}, {Beck}, {Boogert}, {Brown}, {Caselli}, {Charnley}, {Cuppen}, {Dickinson}, {Drozdovskaya}, {Egami}, {Erkal}, {Fraser}, {Garrod}, {Harsono}, {Ioppolo}, {Jim{\'e}nez-Serra}, {Jin}, {J{\o}rgensen}, {Kristensen}, {Lis}, {McCoustra}, {McGuire}, {Melnick}, {{\"O}berg}, {Palumbo}, {Shimonishi}, {Sturm}, {van Dishoeck}, \& {Linnartz}}]{McClure_JWST}
{McClure}, M.~K., {Rocha}, W.~R.~M., {Pontoppidan}, K.~M., {et~al.} 2023, Nature Astronomy

\bibitem[{Mejía {et~al.}(2015)Mejía, {de Barros}, {Seperuelo Duarte}, {da Silveira}, Dartois, Domaracka, Rothard, \& Boduch}]{MEJIA2015}
Mejía, C., {de Barros}, A., {Seperuelo Duarte}, E., {et~al.} 2015, Icarus, 250, 222

\bibitem[{Minissale {et~al.}(2022)Minissale, Aikawa, Bergin, Bertin, Brown, Cazaux, Charnley, Coutens, Cuppen, Guzman, Linnartz, McCoustra, Rimola, Schrauwen, Toubin, Ugliengo, Watanabe, Wakelam, \& Dulieu}]{Minissale_review_22}
Minissale, M., Aikawa, Y., Bergin, E., {et~al.} 2022, ACS Earth and Space Chemistry, 6, 597

\bibitem[{{Mispelaer} {et~al.}(2013){Mispelaer}, {Theul{\'e}}, {Aouididi}, {Noble}, {Duvernay}, {Danger}, {Roubin}, {Morata}, {Hasegawa}, \& {Chiavassa}}]{Mispelaer2013AA}
{Mispelaer}, F., {Theul{\'e}}, P., {Aouididi}, H., {et~al.} 2013, \aap, 555, A13

\bibitem[{{Mousis} {et~al.}(2020){Mousis}, {Aguichine}, {Atkinson}, {Atreya}, {Cavali{\'e}}, {Lunine}, {Mandt}, \& {Ronnet}}]{MousisSSR20}
{Mousis}, O., {Aguichine}, A., {Atkinson}, D.~H., {et~al.} 2020, \ssr, 216, 77

\bibitem[{{Mousis} {et~al.}(2021){Mousis}, {Aguichine}, {Bouquet}, {Lunine}, {Danger}, {Mandt}, \& {Luspay-Kuti}}]{Moussis2021}
{Mousis}, O., {Aguichine}, A., {Bouquet}, A., {et~al.} 2021, The Planetary Science Journal, 2, 72

\bibitem[{{Nanni} {et~al.}(2020){Nanni}, {Burgarella}, {Theul{\'e}}, {C{\^o}t{\'e}}, \& {Hirashita}}]{NanniAA20}
{Nanni}, A., {Burgarella}, D., {Theul{\'e}}, P., {C{\^o}t{\'e}}, B., \& {Hirashita}, H. 2020, \aap, 641, A168

\bibitem[{{Nguyen} {et~al.}(2018){Nguyen}, {Baouche}, {Congiu}, {Diana}, {Pagani}, \& {Dulieu}}]{NguyenAA18}
{Nguyen}, T., {Baouche}, S., {Congiu}, E., {et~al.} 2018, \aap, 619, A111

\bibitem[{{Ninio Greenberg} {et~al.}(2017){Ninio Greenberg}, {Laufer}, \& {Bar-Nun}}]{NinioGreenberg2017}
{Ninio Greenberg}, A., {Laufer}, D., \& {Bar-Nun}, A. 2017, \mnras, 469, S517

\bibitem[{Nixon {et~al.}(2016)Nixon, Pires, Neves, Duque, Jones, Brunger, \& Lopes}]{NIXON201648}
Nixon, K., Pires, W., Neves, R., {et~al.} 2016, International Journal of Mass Spectrometry, 404, 48

\bibitem[{{Noble} {et~al.}(2012){Noble}, {Theule}, {Mispelaer}, {Duvernay}, {Danger}, {Congiu}, {Dulieu}, \& {Chiavassa}}]{NobleAA12}
{Noble}, J.~A., {Theule}, P., {Mispelaer}, F., {et~al.} 2012, \aap, 543, A5

\bibitem[{{Notesco} \& {Bar-Nun}(2000)}]{Notesco2000}
{Notesco}, G. \& {Bar-Nun}, A. 2000, \icarus, 148, 456

\bibitem[{{{\"O}berg} \& {Bergin}(2021)}]{Oberg_Bergin2021}
{{\"O}berg}, K.~I. \& {Bergin}, E.~A. 2021, \physrep, 893, 1

\bibitem[{{{\"O}berg} {et~al.}(2023){{\"O}berg}, {Facchini}, \& {Anderson}}]{Oberg2023ARA}
{{\"O}berg}, K.~I., {Facchini}, S., \& {Anderson}, D.~E. 2023, \araa, 61, 287

\bibitem[{Orient \& Strivastava(1987)}]{orient1987electron}
Orient, O.~J. \& Strivastava, S.~K. 1987, Journal of Physics B: Atomic and Molecular Physics, 20, 3923

\bibitem[{{Poch} {et~al.}(2020){Poch}, {Istiqomah}, {Quirico}, {Beck}, {Schmitt}, {Theul{\'e}}, {Faure}, {Hily-Blant}, {Bonal}, {Raponi}, {Ciarniello}, {Rousseau}, {Potin}, {Brissaud}, {Flandinet}, {Filacchione}, {Pommerol}, {Thomas}, {Kappel}, {Mennella}, {Moroz}, {Vinogradoff}, {Arnold}, {Erard}, {Bockel{\'e}e-Morvan}, {Leyrat}, {Capaccioni}, {De Sanctis}, {Longobardo}, {Mancarella}, {Palomba}, \& {Tosi}}]{PochScience20}
{Poch}, O., {Istiqomah}, I., {Quirico}, E., {et~al.} 2020, Science, 367, aaw7462

\bibitem[{{Potapov} {et~al.}(2020){Potapov}, {J{\"a}ger}, \& {Henning}}]{Potapov2020}
{Potapov}, A., {J{\"a}ger}, C., \& {Henning}, T. 2020, \prl, 124, 221103

\bibitem[{{Redhead}(1962)}]{RedheadVacuum62}
{Redhead}, P.~A. 1962, Vacuum, 12, 203

\bibitem[{{Ruaud} {et~al.}(2016){Ruaud}, {Wakelam}, \& {Hersant}}]{RuaudMNRAS16}
{Ruaud}, M., {Wakelam}, V., \& {Hersant}, F. 2016, \mnras, 459, 3756

\bibitem[{{Rubin} {et~al.}(2023){Rubin}, {Altwegg}, {Berthelier}, {Combi}, {De Keyser}, {Fuselier}, {Gombosi}, {Gudipati}, {H{\"a}nni}, {Kipfer}, {Ligterink}, {M{\"u}ller}, {Shou}, \& {Wampfler}}]{Rubin2023}
{Rubin}, M., {Altwegg}, K., {Berthelier}, J.-J., {et~al.} 2023, \mnras, 526, 4209

\bibitem[{{Sandford} \& {Allamandola}(1988)}]{Sandford_Allamandola_icarus_88}
{Sandford}, S.~A. \& {Allamandola}, L.~J. 1988, \icarus, 76, 201

\bibitem[{{Schmitt}(1992)}]{Schmitt92}
{Schmitt}, B. 1992, in Interrelations Between Physics and Dynamics for Minor Bodies in the Solar System, ed. D.~{Benest} \& C.~{Froeschle}, 265

\bibitem[{{Simon} {et~al.}(2019){Simon}, {{\"O}berg}, {Rajappan}, \& {Maksiutenko}}]{Simon2019}
{Simon}, A., {{\"O}berg}, K.~I., {Rajappan}, M., \& {Maksiutenko}, P. 2019, \apj, 883, 21

\bibitem[{{Simon} {et~al.}(2023){Simon}, {Rajappan}, \& {{\"O}berg}}]{Simon2023}
{Simon}, A., {Rajappan}, M., \& {{\"O}berg}, K.~I. 2023, \apj, 955, 5

\bibitem[{Smith {et~al.}(1997)Smith, Huang, Wong, \& Kay}]{KayPRL97}
Smith, R.~S., Huang, C., Wong, E. K.~L., \& Kay, B.~D. 1997, Phys. Rev. Lett., 79, 909

\bibitem[{Smith {et~al.}(2011)Smith, Matthiesen, Knox, \& Kay}]{smith_JPCA_2011}
Smith, R.~S., Matthiesen, J., Knox, J., \& Kay, B.~D. 2011, The Journal of Physical Chemistry A, 115, 5908, pMID: 21218834

\bibitem[{Souda(2007)}]{SoudaPhysRevB}
Souda, R. 2007, Phys. Rev. B, 75, 184116

\bibitem[{{Stevenson} {et~al.}(1999){Stevenson}, {Kimmel}, {Dohnalek}, {Smith}, \& {Kay}}]{Stevenson1999}
{Stevenson}, K.~P., {Kimmel}, G.~A., {Dohnalek}, Z., {Smith}, R.~S., \& {Kay}, B.~D. 1999, Science, 283, 1505

\bibitem[{Straub(1995)}]{straub1995absolute}
Straub, H. 1995, Physical Review A, 52, 1115

\bibitem[{{Sutherland} {et~al.}(2018){Sutherland}, {Dopita}, {Binette}, \& {Groves}}]{SutherlandASCL18}
{Sutherland}, R., {Dopita}, M., {Binette}, L., \& {Groves}, B. 2018, {MAPPINGS V: Astrophysical plasma modeling code}

\bibitem[{{Taquet} {et~al.}(2012){Taquet}, {Ceccarelli}, \& {Kahane}}]{TaquetAA12}
{Taquet}, V., {Ceccarelli}, C., \& {Kahane}, C. 2012, \aap, 538, A42

\bibitem[{{Theul{\'e}}(2020)}]{Theule2020IAUS}
{Theul{\'e}}, P. 2020, in Laboratory Astrophysics: From Observations to Interpretation, ed. F.~{Salama} \& H.~{Linnartz}, Vol. 350, 139--143

\bibitem[{{Theul{\'e}} {et~al.}(2013){Theul{\'e}}, {Duvernay}, {Danger}, {Borget}, {Bossa}, {Vinogradoff}, {Mispelaer}, \& {Chiavassa}}]{TheuleASR2013}
{Theul{\'e}}, P., {Duvernay}, F., {Danger}, G., {et~al.} 2013, Advances in Space Research, 52, 1567

\bibitem[{{Tielens} \& {Hagen}(1982)}]{TielensAA82}
{Tielens}, A.~G.~G.~M. \& {Hagen}, W. 1982, \aap, 114, 245

\bibitem[{{Tinacci} {et~al.}(2022){Tinacci}, {Germain}, {Pantaleone}, {Ferrero}, {Ceccarelli}, \& {Ugliengo}}]{Tinacci2022}
{Tinacci}, L., {Germain}, A., {Pantaleone}, S., {et~al.} 2022, ACS Earth and Space Chemistry, 6, 1514

\bibitem[{{Tobin} {et~al.}(2023){Tobin}, {van't Hoff}, {Leemker}, {van Dishoeck}, {Paneque-Carre{\~n}o}, {Furuya}, {Harsono}, {Persson}, {Cleeves}, {Sheehan}, \& {Cieza}}]{Trobin2023Natur}
{Tobin}, J.~J., {van't Hoff}, M. L.~R., {Leemker}, M., {et~al.} 2023, \nat, 615, 227

\bibitem[{Tonauer {et~al.}(2023)Tonauer, Fidler, Giebelmann, Yamashita, \& Loerting}]{Tonauer2023}
Tonauer, C.~M., Fidler, L.-R., Giebelmann, J., Yamashita, K., \& Loerting, T. 2023, Journal of Chemical Physics, 158, 141001

\bibitem[{Ulbricht {et~al.}(2006)Ulbricht, Zacharia, Cindir, \& Hertel}]{Ulbricht06}
Ulbricht, H., Zacharia, R., Cindir, N., \& Hertel, T. 2006, CARBON, 44, 2931

\bibitem[{{Vasyunin} \& {Herbst}(2013)}]{VasyuninApJ13}
{Vasyunin}, A.~I. \& {Herbst}, E. 2013, \apj, 762, 86

\bibitem[{Villanueva {et~al.}(2011)Villanueva, Mumma, DiSanti, Bonev, Gibb, Magee-Sauer, Blake, \& Salyk}]{Villanueva2011}
Villanueva, G., Mumma, M., DiSanti, M., {et~al.} 2011, Icarus, 216, 227

\bibitem[{{Viti} {et~al.}(2004){Viti}, {Collings}, {Dever}, {McCoustra}, \& {Williams}}]{Viti2004MNRAS}
{Viti}, S., {Collings}, M.~P., {Dever}, J.~W., {McCoustra}, M. R.~S., \& {Williams}, D.~A. 2004, \mnras, 354, 1141

\bibitem[{Zawadzki(2018)}]{Zawadzki2018}
Zawadzki, M. 2018, European Physical Journal D, 72, 12

\end{thebibliography}

\appendix
\section{Surface coverage calibration}
\label{Appendix}

\begin{figure}[ht]
   \centering
   \includegraphics[width = 9 cm]{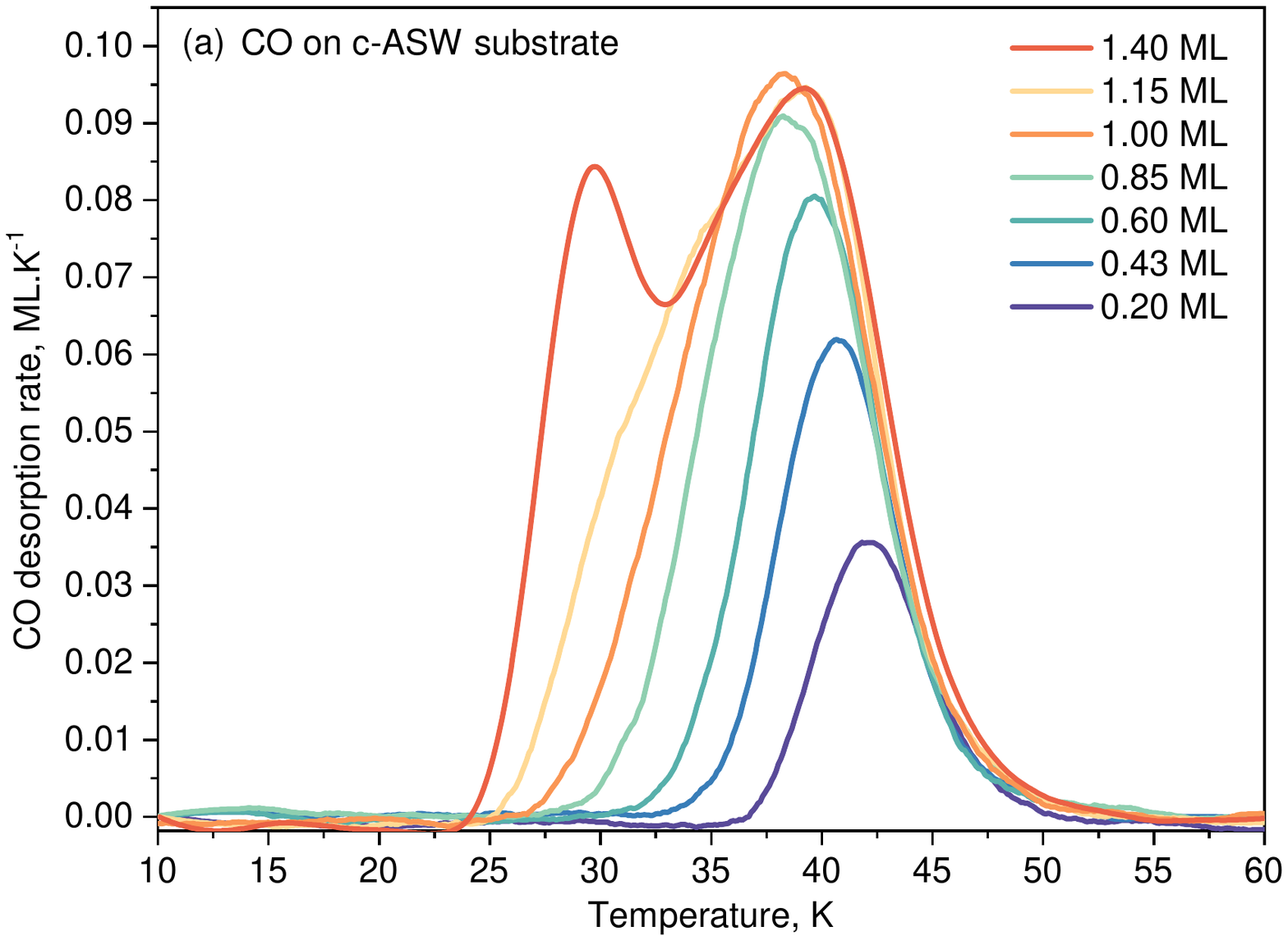}
    \caption{TPD curves for varying CO coverages on c-ASW, highlighting the multi-layer desorption peak near 25 K indicative of 1 ML coverage.}
   \label{CO_familly}
    \end{figure}

To calibrate surface coverages, CO is incrementally deposited on compact amorphous solid water (c-ASW) to identify the dose corresponding to 1 ML based on the observed `filling behaviour' during desorption, as characterised by \cite{KimmelJChPh2001_exp}. TPD studies indicate that CO desorption from submonolayer coverage occurs over a broad temperature range (25K to 55K), with the highest energy sites releasing CO at higher temperatures. In Fig. \ref{CO_familly}, each curve corresponds to a given dose of CO, that is the initial coverage at 10 K. As coverage increases, desorption shifts to lower temperatures due to the occupation of lower energy sites, causing the TPD peak to shift accordingly. This is consistently demonstrated by the leading edge of the TPD profiles moving to lower temperatures with greater CO coverage, while the tail of high-binding-energy sites remains unchanged. Notably, the emergence of a low-temperature peak past 1 ML coverage marks the transition to multilayer desorption, following zeroth-order kinetics.

\end{document}